\documentclass[prl,reprint,aps,amsmath,amssymb,amsfonts,superscriptaddress,notitlepage]{revtex4-1}

\usepackage[colorlinks=true,citecolor=blue,urlcolor=blue,linkcolor=blue,hyperfigures=true]{hyperref}
\usepackage{graphicx}
\usepackage{amsmath,amsfonts}
\usepackage[dvipsnames]{xcolor}
\usepackage{color}

\usepackage{inputenc}
\usepackage{comment}

\newcommand{\up}{\uparrow}
\newcommand{\dn}{\downarrow}

\newcommand{\kv}{\ensuremath{\mathbf{k}}}
\newcommand{\qv}{\ensuremath{\mathbf{q}}}

\newcommand{\av}[1]{\ensuremath{\left\langle #1 \right\rangle}}

\newcommand{\abs}[1]{\ensuremath{\left| #1 \right|}}

\renewcommand{\Re}{\operatorname{Re}}
\renewcommand{\Im}{\operatorname{Im}}

\definecolor{ErikBlauw}{cmyk}{1.,0.3,0.,0.} 
\definecolor{ErikRood}{cmyk}{.2,0.9,0.8,0.} 
\definecolor{ErikGroen}{cmyk}{1,0.20,1,0} 

\definecolor{ErikOranje}{HTML}{F68542}
\definecolor{ErikGrijs}{HTML}{868586}
\definecolor{ErikGrijs2}{HTML}{AAAAAA}
\definecolor{ErikWit}{HTML}{FFFFFF}
\definecolor{ErikZwart}{HTML}{000000}

\definecolor{erik_red}{RGB}{215,25,28}
\definecolor{erik_redb}{RGB}{255,113,110}
\definecolor{erik_blueb}{RGB}{5,113,176}
\definecolor{erik_yellow}{RGB}{255,255,191}
\definecolor{erik_white}{RGB}{255, 255, 255}

\definecolor{colorRed}{RGB}{228,26,28}
\definecolor{colorBlue}{RGB}{55,126,184}
\definecolor{colorGreen}{RGB}{77,175,74}
\definecolor{colorPurple}{RGB}{152,78,163}
\definecolor{colorOrange}{RGB}{255,127,0}

\def\presuper#1#2%
  {\mathop{}%
   \mathopen{\vphantom{#2}}^{#1}%
   \kern-0.5\scriptspace%
   #2}

\usepackage{tikz}
\usetikzlibrary{calc}
\usetikzlibrary{arrows}
\usetikzlibrary{patterns}
\usetikzlibrary{decorations.pathmorphing}
\usetikzlibrary{decorations.pathreplacing}
\usetikzlibrary{decorations.markings}
\usetikzlibrary{shapes.geometric}
\usetikzlibrary{positioning}

\begin{document}

 \author{Erik G. C. P. van Loon}
 \email{evloon@itp.uni-bremen.de}
 \affiliation{Institut f{\"u}r Theoretische Physik, Universit{\"a}t Bremen, Otto-Hahn-Allee 1, 28359 Bremen, Germany}
 \affiliation{Bremen Center for Computational Materials Science, Universit{\"a}t Bremen, Am Fallturm 1a, 28359 Bremen, Germany}

\author{Friedrich Krien}
\affiliation{Jo\v{z}ef Stefan Institute, Jamova 39, SI-1000, Ljubljana, Slovenia}

\author{Andrey A. Katanin}
\affiliation{Moscow Institute of Physics and Technology, 141701 Dolgoprudny, Russia}
\affiliation{M. N. Mikheev Institute of Metal Physics, Russian Academy of Sciences, 620108 Yekaterinburg, Russia}

\title{The Bethe-Salpeter equation at the critical end-point of the Mott transition }

\begin{abstract}
Strong repulsive interactions between electrons can lead to a Mott metal-insulator transition. 
The Dynamical Mean-Field Theory (DMFT) explains the critical end-point and the hysteresis region\, usually in terms of single-particle concepts
such as the spectral function and the quasiparticle weight. 
In this work, we reconsider the critical end point of the metal-insulator transition on DMFT's two-particle level.
We show that the relevant eigenvalue and eigenvector of the non-local Bethe-Salpeter kernel in the charge channel provide a unified picture of the hysteresis region and of the critical end point of the Mott transition. In particular, they simultaneously explain the thermodynamics of the hysteresis region and the iterative stability of the DMFT equations.
This analysis paves the way for a deeper understanding of phase transitions in correlated materials.
\end{abstract}

\maketitle

\begin{figure}
\includegraphics{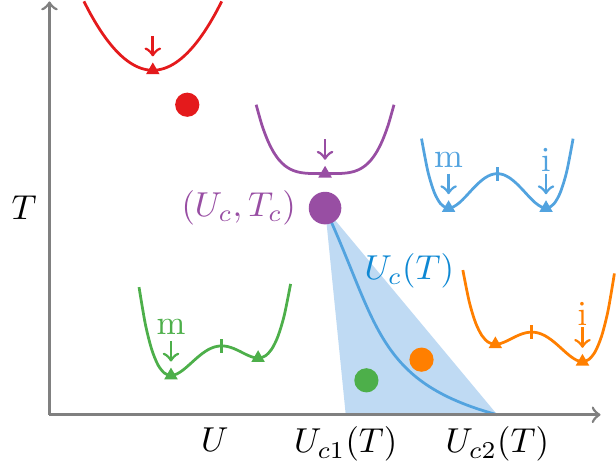}
\caption{Sketch of the phase diagram of the particle-hole symmetric Hubbard model in DMFT. The first-order metal-insulator transition occurs at $U_c(T)$ (blue curve), with a second-order critical end-point at $(U_c,T_c)$ (purple dot). The shaded area is the hysteresis region, where both metal (m) and insulator (i) can be stabilized. The colored curves illustrate the free energy landscape at selected points (dots) in the phase diagram, the vertical marks denote the local maxima of the free energy, the triangles the local minima and the arrows the global minimum. }
\label{fig:sketch}
\end{figure}

The interplay of interactions, correlations and quantum statistics in quantum many-body physics is responsible for the appearance of complicated new phases, with the Mott transition~\cite{Imada1998} as a prominent example. 
The simplest theoretical realization of this correlation driven metal-insulator transition (MIT) occurs in the (single-band) Hubbard model~\cite{Hubbard63,Kanamori63,Gutzwiller63,Hubbard64}.
Quantum simulators using ultracold fermions in optical lattices are providing unprecedented experimental insight into this transition~\cite{Jordens08,Schneider08,Jordens10,Duarte15,Greif16}.

 From the theory side, the Dynamical Mean-Field theory~\cite{Metzner89,Georges96} (DMFT) provides a rare example of an exact solution to a strongly correlated problem, namely to the Hubbard model in the limit of infinite dimensions. 
 During the first decade after DMFT's invention, the essence~\footnote{New perspectives still appear, such as topological views on the transition~\cite{Logan15,Sen20}.} of the Mott transition was ascertained~\cite{Jarrell92,Georges92,Georges93,Zhang93,Rozenberg94,Noack99,Bulla99,Blumerphd}: At the zero temperature transition to the insulating phase, the quasiparticle weight vanishes and the self-energy is divergent at small frequency, in contrast to the Fermi liquid. 
The $U$-$T$ (interaction-temperature) DMFT phase diagram of the particle-hole symmetric model can be summarized as follows (sketched in Fig.~\ref{fig:sketch}, for an overview see Refs.~\cite{Blumerphd,Eckstein07,Strand11,Schafer15}): at low temperature, there is a metallic phase at small $U<U_{c1}$ and an insulating phase at large $U>U_{c2}$. In between, for $U_{c1} < U <U_{c2}$, both metallic and insulating solutions can be stabilized. This hysteresis region (shaded blue area) ends at a critical temperature $T_c$, where $U_{c1}=U_{c2}=U_{c}$ (purple dot).
 No phase separation occurs in the particle-hole symmetric system~\cite{Eckstein07}.

Although the single-particle properties (Green's function, self-energy, quasiparticle weight) are sufficient to understand the essentials of the metal-insulator transition, two-particle properties provide another rich layer of information about the response to external fields, spatial correlations, and optical properties.
The simplifications of infinite dimensions allowed early studies at the two-particle level~\cite{Khurana90,Zlatic90,Pruschke93,Zhang93,Rozenberg94,Rozenberg95}, but a systematic investigation of the DMFT two-particle physics had to wait~\cite{Brener08,Rohringer12,Boehnke12,Rohringer12,vanLoon14b,Geffroy19,Strand19,Krien19,Melnick20}
for computational improvement, especially the invention of continuous-time Quantum Monte Carlo solvers~\cite{Rubtsov05,Werner06,Gull11}.

There has recently been a flurry of activity on divergences on the two-particle level~\cite{Schafer13,Kozik15,Schafer16,Gunnarsson17,Melnick20,Chalupa20}, from simple toy models~\cite{Stan15,Rossi15} and the Hubbard atom~\cite{Thunstrom18} to cluster approaches~\cite{Vucicevic18}, relating these divergences to unphysical solutions~\cite{Kozik15,Gunnarsson17,Tarantino17} and to the suppression of fluctuations~\cite{Chalupa18,Springer19}.
Crucially, divergences of the irreducible vertex already appear in impurity models and therefore cannot originate
in the Mott transition:
there is no Mott transition in an impurity model with fixed bath -- just as the Brillouin function in Curie-Weiss mean field theory of the Ising model is smooth -- and only the self-consistent adjustment of the DMFT auxiliary impurity provides the opportunity for a phase transition. Thus, on the two-particle level we also expect the Mott transition to appear via self-consistent feedback, that is, outside the impurity model. 

The divergences of the irreducible vertex imply that the eigenvalues of the local charge vertex function and local generalized susceptibility can change sign~\cite{Gunnarsson17,Springer19,Melnick20} and,
as a matter of fact, the same holds for the corresponding lattice quantities.
This undercuts the original idea of using them for constructing the Landau functional near the Mott transition ~\cite{Chitra01,Potthoff03}
because the curvature of the free energy is supposed to be positive definite for stationary solutions.
Indeed, Ref.~\cite{Potthoff03} pointed out that the stationary point of the self-energy functional
is not necessarily an extremum.
Recently, it was shown \cite{Krien19} that the \textit{non-local} Bethe-Salpeter kernel,
instead of the full one, is a more appropriate quantity to describe the Mott transition,
since it yields {positive} eigenvalues which approach unity from below.  
The corresponding symmetric Landau parameter is indeed not affected by the divergences of the irreducible vertex~\cite{Krien19,Melnick20}. 

We show here that the non-local Bethe-Salpeter kernel, associated with the charge sector, provides an intriguing new view on the Mott transition
across the hysteresis region and especially at the critical end point.
In particular, it appears in the expression for the second derivative of
an appropriate Landau functional for the Mott transition, yielding a positive curvature for stationary solutions,
whereas the functionals of Refs.~\cite{Chitra01,Potthoff03} should be used at weak coupling. 
Furthermore, this kernel is directly related to the Jacobian of the DMFT fixed point function~\cite{Blumerphd,Zitko09,Strand11}, which determines the stability of iterative solutions. 
The leading eigenvalue of the kernel is unity at the finite temperature critical end point,
signalling the onset of the hysteresis region.
Nevertheless, at particle-hole symmetry the frequency structure of the corresponding eigenvector ensures that the compressibility does not diverge, 
cf. Ref. ~\cite{Reitner20}.
Therefore, the non-local Bethe-Salpeter kernel determines two apparently separate stability criteria, the thermodynamic and the iterative stability, and the eigenvector frequency structure -- given by the difference between insulating and metallic solution -- distinguishes between diverging response and exact cancellation.

We consider the Hubbard model describing the competition between localization due to the Coulomb interaction $U$ and delocalization due to the dispersion $t_{\bf k}$.
We use $i$ to label the sites on the periodic lattice and $\kv$ to label the corresponding momentum. 
The model is given by the Hamiltonian
 \begin{align}
     H = -\sum_{\kv,\sigma} t_{\kv} c^\dagger_{\kv\sigma} c^{\phantom{\dagger}}_{\kv\sigma} + U \sum_i n_{i\up}n_{i\dn},
 \end{align}
where $c^\dagger_{\kv\sigma}$ is the creation operator for a fermion with momentum $\kv$ and spin $\sigma =\up,\dn$ and $n_{i\sigma}=c^\dagger_{i\sigma}c^{\phantom{\dagger}}_{i\sigma}$ is the number operator of electrons with spin $\sigma$ on site $i$.  We consider this model in the grand-canonical ensemble at temperature $T$ and chemical potential $\mu$. A central object of interest is the (one-particle) Green's function $G_{\kv,\nu,\sigma}=-\av{c_\sigma c_\sigma^\dagger}_{\kv,\nu}$ in the Matsubara formalism, where $\nu_n = \pi T (2n+1)$, with $n\in\mathbb{Z}$ the fermionic Matsubara frequencies.  
We consider the paramagnetic state and for compactness drop the spin labels.

The Dynamical Mean-Field Theory~\cite{Metzner89,Georges96} (DMFT) provides an approximate solution to this model by setting $\Sigma_{\kv,\nu}=\Sigma^\text{AIM}_{\nu}$, where AIM stands for an auxiliary impurity model consisting of a single interacting site in a self-consistently determined bath.
For the present discussion, it is sufficient to state that the auxiliary impurity model
serves as a tool to evaluate the functional relation $\Sigma[\Delta]$ between the bath hybridization function $\Delta$ and the self-energy $\Sigma$ of the AIM
(in practice, we use the ALPS~\cite{ALPS2} and iQIST \cite{Huang15,Huang17}
realizations of CTQMC~\cite{Gull11} solver of Ref.~\cite{Hafermann13} with improved estimators~\cite{Hafermann12}). 
The hybridization $\Delta$ of the auxiliary impurity model is chosen so that the mean-field self-consistency equation  $g_\nu[\Delta]=f(\Delta_\nu,g_\nu[\Delta])$ is satisfied. Here
$g_\nu[\Delta]=1/(i \nu_n-\Delta_\nu-\Sigma_\nu[\Delta])$ and 
\begin{align}
 f(\Delta_\nu,g_\nu)&= \sum_\kv G_{\kv,\nu} = \sum_\kv \frac{1}{
 g_\nu^{-1}+\Delta_\nu+t_{\kv}
 },
 \label{eq:fix0}
\end{align}
from now on $\sum_\kv\equiv\frac{1}{N} \sum_{\kv \in \text{BZ}}$ denotes the momentum average over the Brillouin Zone. The square brackets denote functional relations.

In this work, we consider the two-dimensional square lattice Hubbard model, $t_\kv=2t (\cos k_x + \cos k_y)$ at half-filling. The energy scale is set by $4t=1$. The half-filled model is particle-hole symmetric, which leads to $\Re g_\nu=0$ and $\Re \Sigma_\nu =U/2$. In other words, only the imaginary parts of both quantities are of interest, which simplifies the analysis.
 
\emph{Fixed point equation: }%
The auxiliary impurity model is a finite system that cannot undergo a (finite temperature) phase transition by itself. 
Instead, as in Weiss' mean field theory of magnetism, it is the self-consistency condition that opens the possibility of a phase transition. Therefore, our analysis of the critical point starts with the self-consistency condition.

DMFT looks for solutions of Eq.~\eqref{eq:fix0}, i.e., a fixed point $\Delta^\ast=h[\Delta^\ast]$ where $h[\Delta]=i\nu_n-\Sigma_\nu[\Delta]-1/f(\Delta_{\nu},g_{\nu}[\Delta])$.
To avoid issues related to the non-invertibility~\cite{Kozik15} of the mapping $\Delta\mapsto g$, we perform the stability analysis in terms of the iterative scheme $\Delta^{(n+1)} = h[\Delta^{(n)}]$.
An important question is if these iterations converge to the fixed point $\Delta^\ast$ 
if one starts the iteration close to $\Delta^\ast$. In that case, the fixed point is called attractive~\footnote{Here we ignore the possibility of mixing of previous and current iterative solutions, since we are interested in the fundamental aspects of the stability of DMFT solutions.}. 
The textbook analysis, based on a linear expansion of $h$ around the fixed point, shows that $\Delta^\ast$ is attractive if and only if all eigenvalues of the Jacobian $\mathcal{J}|_{\Delta^\ast}=(\delta h/\delta \Delta)|_{\Delta^\ast}$ have magnitude smaller than 1. 
Any eigenvalue larger than 1 implies that the self-consistency cycle is repulsive along the direction given by the corresponding eigenvector.
For DMFT, the Jacobian can be evaluated explicitly in Matsubara space as (see Supplementary Material)
\begin{align}
\hat{\mathcal{J}}_{\nu \nu ^{\prime }} &= \hat{x}^{-1}\hat{\mathcal{D}}_{\nu\nu ^{\prime }}\hat{x}  \label{Jak} \\
\mathcal{D}_{\nu \nu ^{\prime }} &=T \left( \sum\limits_{\mathbf{k}}G_{\mathbf{k},\nu}^{2}-g_{\nu
}^{2}\right) F^{\rm loc}_{\omega=0,\nu \nu ^{\prime }},\label{DBSK} 
\end{align}%
where 
$F^{\rm loc}_{\omega,\nu \nu ^{\prime }}$ is the full local charge vertex, $\hat{x}_{\nu\nu'}=-T \delta_{\nu\nu'} g_\nu g_\nu$ is the local ``bubble''. The hat denotes a matrix in Matsubara space and, when possible, the matrix indices $\nu$, $\nu'$ are dropped.
The essential element of Eq.~\eqref{Jak} is the non-local Bethe-Salpeter kernel $\mathcal{D}$ at $\qv=0$ and $\omega=0$ --- a quantity that also appears in the calculation of linear response functions based on a decomposition into local and non-local fluctuations.

 \emph{Response functions: }%
 Indeed, the DMFT recipe provided above not only allows us to determine the one-particle Green's function $G$ for a given set of parameters $(U,\mu,T)$. On top of this, DMFT also describes how the system would (linearly) respond~\cite{Georges96} to an external field with frequency $\omega$ and momentum $\qv$. We restrict our analysis to time-independent fields, $\omega=0$. The response function $\chi_{\qv=0}$ can be
 obtained from 
 \begin{align}  \hat{\chi}^{\text{DMFT}}_{\qv =0}
 =
 \left(\hat{1}-\hat{x} \hat{F}\right)
 \frac{\hat{1}}{\hat{1}-\hat{\mathcal{D}}} \hat{X}_{\qv=0}, \label{eq:BSE}
\end{align}
where $\left(\hat{X}_\qv\right)_{\nu\nu'}=-T \delta_{\nu\nu'}\sum_\kv G_{\kv,\nu}G_{\kv+\qv,\nu}$ is the full bubble, 
the fraction denotes matrix inversion in Matsubara space. 
The relation~\eqref{eq:BSE}, which is derived in the Supplemental Material, is a resummation~\cite{Rubtsov08,Brener08,Hafermannphd,Rohringer18} of the more familiar expression~\cite{Georges96} $\hat{\chi}=(1+\hat{X}\hat{\Gamma})^{-1}\hat{X}$ that avoids the divergences of the irreducible vertex $\Gamma$. From this generalized susceptibility matrix, the physical response function is obtained as a sum over both fermionic frequencies.
For example, the compressibility $dn/d\mu$ is obtained from the generalized susceptibility at $\qv=0$ (and, as before, $\omega=0$), 
\begin{align}
    \frac{dn}{d\mu} =& \sum_{\nu\nu'} \left(\hat{\chi}^\text{DMFT}_{\qv=0}\right)_{\nu\nu'}. \label{eq:dndmutrace}
\end{align}
The response in DMFT is thermodynamically consistent in the sense that this Bethe-Salpeter determination of $dn/d\mu$ gives the same result as changing $\mu$ explicitly and calculating the change in $n$~\cite{vanLoon15}. 

\emph{Landau theory: }%
Following Landau, the free energy functional is the essential ingredient for understanding stable and unstable phases and hysteresis close to the critical point. Characteristic free energy curves are sketched in Fig.~\ref{fig:sketch}.
The second derivative of the free energy determines if the stationary point is a local minimum ($\delta^2 F>0$, stable, denoted by triangles in Fig.~\ref{fig:sketch}) or a local maximum ($\delta^2 F<0$, unstable, denoted by a vertical bar).
The critical point is where a stable point turns unstable, in other words, $\delta^2 F=0$ exactly at the critical point (purple curve in Fig.~\ref{fig:sketch}).

The Mott transition on the Bethe lattice has been studied using Landau theory~\cite{Kotliar99,Kotliar00,Blumerphd}. Here we generalize this approach to arbitrary dispersion $t_\kv$. With the hybridization $\Delta$ as the order parameter, we write the Landau functional $\Omega$ as $\Omega[\Delta]= \Omega_\text{imp}[\Delta]-\Omega'[\Delta]$ (see Supplementary material for more details), where {$\Omega_\text{imp}$} is the thermodynamic potential of the auxiliary impurity model. $\Omega'$ provides the non-local feedback and ensures that the first derivative $\delta \Omega/\delta \Delta=0$ at the self-consistent DMFT solution, that is, 
\begin{align}
\frac{\delta \Omega}{\delta \Delta_\nu}=T\left(g_\nu[\Delta]-g^{\rm sc}_\nu[\Delta]\right)\label{Omega1},
\end{align}
where $g_\nu[\Delta]=\delta \Omega_\text{imp}[\Delta]/\delta \Delta_\nu$ is the local Green's function determined from the AIM for a given hybridization $\Delta$  and $g^\text{sc}[\Delta]=\delta \Omega'/\delta \Delta$ is the solution of the self-consistency condition $g_\nu^{\rm sc}=f(\Delta_\nu,g^{\rm sc}_\nu[\Delta])$  for a given $\Delta$, see Eq.~\eqref{eq:fix0}. We note that the map $\Delta\mapsto g^{\rm sc}$ can be multivalued. However, as we argue in the Supplementary Material, at sufficiently strong coupling (e.g., near the MIT), only one branch is relevant.

To determine the stability of this solution, we proceed with the second derivative (Hessian), which reads (see Supplementary Material)
\begin{align}
    \frac{\delta^2 \Omega}{\delta \Delta_\nu \delta \Delta_{\nu'}}=
        -\frac{\hat{1}}{\hat{1}-\hat{x}^{-1}\hat{X}}
    \left(\hat{1}-\hat{\mathcal{D}}\right)\hat{x}.
    \label{eq:OmegaStable}
\end{align}
This is a matrix equation in Matsubara space, $\delta^2 \Omega/\delta (i\Delta)^2$ is the Hessian matrix, which is a real matrix in the case of particle-hole symmetry. The factor $(\hat{1}-\hat{x}^{-1} \hat{X})^{-1}$ is diagonal in frequency, and, as we discuss in the Supplementary Material, in the entire region of interest it has only positive elements, therefore the stability is determined by eigenvalues of $\hat{1}-\hat{\mathcal{D}}$.
At the critical point, the Hessian changes from stable to unstable, i.e., one eigenvalue of $\delta^2 \Omega /\delta \Delta^2$ is equal to zero, which requires an eigenvalue of unity for $\hat{\mathcal{D}}$.

\begin{figure}
    \centering
    \includegraphics[width=0.85\linewidth]{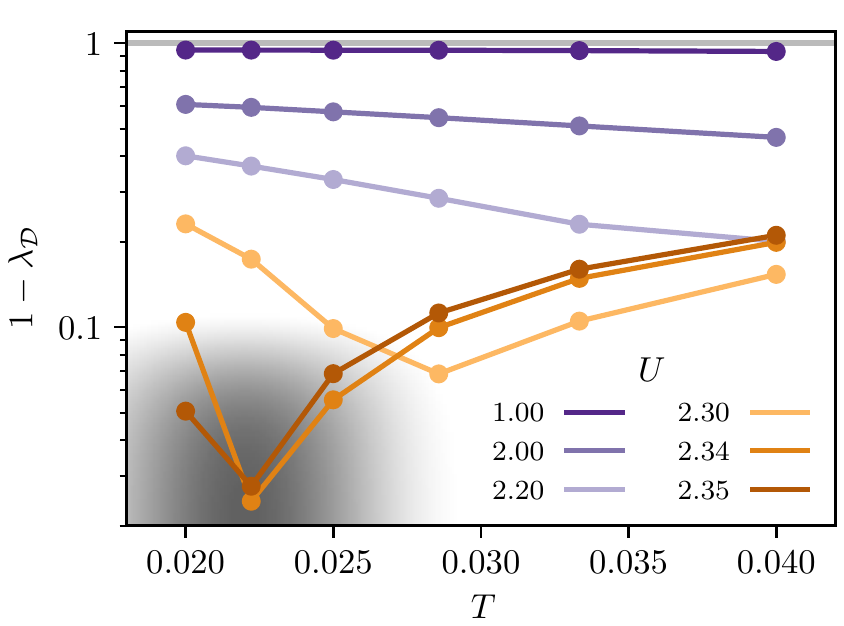}
    \caption{
The leading eigenvalue (note the logarithmic scale) of $\hat{\mathcal{D}}$ approaches unity close to the critical point, $2.3<U_c<2.35$ and $0.02<T_c<0.025$ (gray region).}
    \label{fig:ev}
\end{figure}

The same non-local Bethe-Salpeter kernel $\hat{\mathcal{D}}$ has appeared three times in stability criteria:
in the Jacobian of the fixed point equation; in the compressibility;  and in the second derivative of the self-energy functional. 
The latter two relate to the \emph{stability} of the physical solution,
whereas the Jacobian determines the \emph{attractiveness} of the fixed point in an iterative scheme.
For DMFT, these two aspects are tied together by a single kernel.

This allows us to create a unified picture of the hysteresis region  of the particle-hole symmetric metal-insulator transition. 
At the critical end point $(U_c,T_c)$, the purple dot in Fig.~\ref{fig:sketch}, the two stable (triangles in Fig.~\ref{fig:sketch}) and the one unstable (vertical marks in Fig.~\ref{fig:sketch}) stationary points merge together.
Therefore, the quadratic part of the free energy functional vanishes at this point (purple curve), which together with Eq.~\eqref{eq:OmegaStable}  means that the Bethe-Salpeter kernel $\mathcal{D}$ has an eigenvector $V$ with eigenvalue $\lambda \rightarrow 1$ (Fig.~\ref{fig:ev}) exactly at the critical end point.
Since $\hat{\mathcal{D}}$ is related to the Jacobian of the fixed point equation, the stable and unstable solutions correspond to attractive and repulsive fixed points, respectively~\cite{Strand11}. 

\begin{figure}
    \centering
    \includegraphics{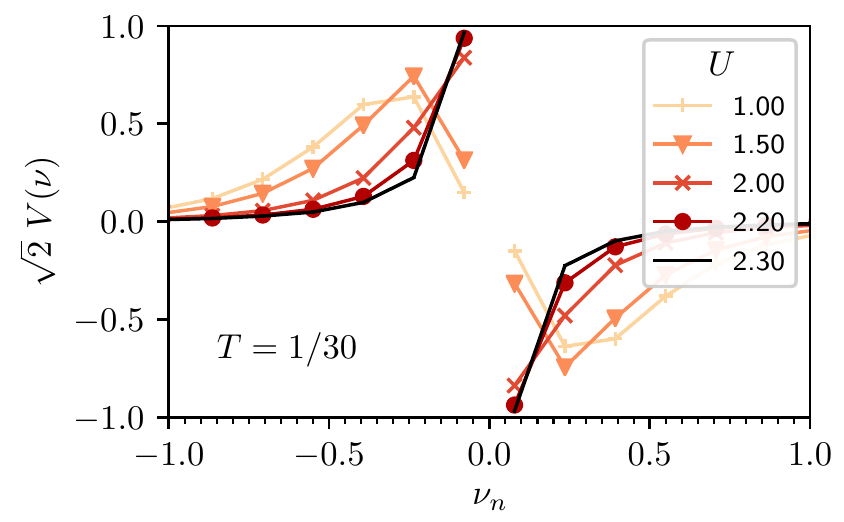}
    \caption{The leading right eigenvector $V$ of the non-local Bethe-Salpeter kernel $\hat{\mathcal{D}}$, for $T$ just above $T_c$. As $U$ increases and the Mott transition is approached, the eigenvector localizes around $\nu = 0$ and $\lambda\rightarrow 1$. The eigenvector is normalized to $\sum_\nu \abs{V(\nu)}^2=1$.} 
    \label{fig:eigenvector}
\end{figure}

\begin{figure}
    \centering
    \includegraphics{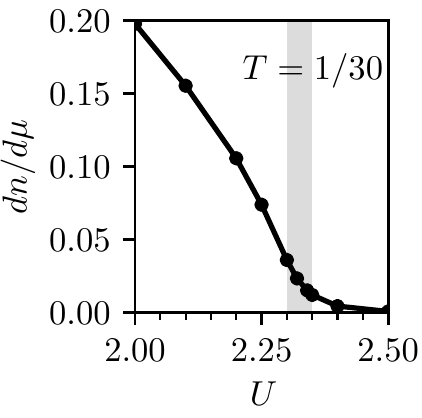}%
    \includegraphics{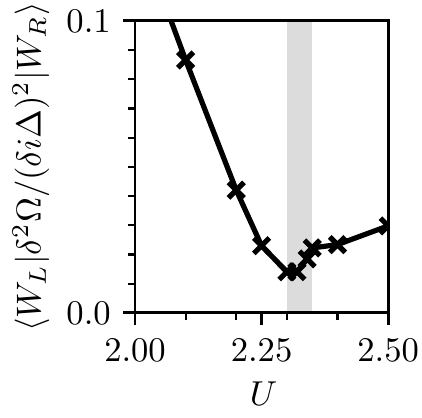}
    \caption{Left: The compressibility does not diverge as the Mott transition is approached. 
    Right: The second derivative of the Landau functional $\Omega$ determines the thermodynamic stability. 
    Approaching the Mott critical point, the vanishing of this second derivative (in the direction given by $W_R$) signals the onset of the phase transition. Here, $W_{L,R}=g_\nu^{\pm 2} V$ where $V$ is the leading right eigenvector of the non-local Bethe-Salpeter kernel (cf.  Fig.~\ref{fig:eigenvector}), and $T=1/30 >T_c$. The gray band indicates the vicinity of the Mott transition, $2.3<U_c<2.35$. 
}
    \label{fig:obs}
\end{figure}

Figure~\ref{fig:eigenvector} shows the leading right eigenvector $V$ of $\hat{\mathcal{D}}$ close to the critical end point.
The physical meaning of this eigenvector is that it relates the three fixed points that exist at $T<T_c$, as $\Delta_m(\nu)-\Delta_u(\nu) \propto (T_c-T)^\beta
V(\nu)$ 
and $\Delta_i(\nu)-\Delta_u(\nu) \propto (T_c-T)^\beta 
V(\nu)$, where $\Delta_m$, $\Delta_i$ and $\Delta_u$ are the hybridization functions at the metallic, insulating and unstable fixed points, respectively, and $\beta$ is a critical exponent. This together with particle-hole symmetry [$\Delta(\nu)=-\Delta(-\nu)$] implies $V(\nu)=-V(-\nu)$, i.e., the eigenvector $V$ is antisymmetric~\cite{Springer19}.
As the difference between solutions, $V$ provides the ``order parameter'' -- similar to Kotliar's~\cite{Kotliar99} $\delta \Delta_L$ at $T=0$ -- in the sense of Landau's functional: At the critical point, the free energy landscape goes from a parabola to a double well potential along the direction given by $V$.
Figure~\ref{fig:obs} shows that the second derivative of the grand potential -- along the direction given by the right eigenvector $W_R$ of the Jacobian -- indeed goes to zero as one gets close to the Mott critical end point. Note that the figure is at $T>T_c$, so the second derivative does not quite reach zero.

Since $V(\nu) \sim \delta_{\nu,\nu_0}-\delta_{\nu,-\nu_0}$ at the critical point (cf. Fig.~\ref{fig:eigenvector}), the three solutions $\Delta(\nu)$ differ only at low frequency, i.e., close to the Fermi level. This is in agreement with what is known qualitatively from investigations of the Density of States: the difference between the insulator and the metal is that the latter has a quasiparticle peak at the Fermi level.
Astretsov et al.~\cite{Astretsov20} used a single Matsubara frequency approximation to study the cuprates, our result here is a direct quantitative proof that this kind of approximation is justified at the critical end point of the Mott transition. 

At $T<T_c$ and $U_{c1}<U<U_{c2}$, the Bethe-Salpeter equation is convergent (and the iterative scheme is attractive) at both the metallic and the insulating solutions, $\lambda<1$, and divergent (repulsive) at the unstable fixed point, $\lambda>1$. Although both metallic and insulating solution are attractive fixed points, only one of them is the global minimum (c.f., green and orange curves in Fig.~\ref{fig:sketch}) of the free energy in most of the hysteresis region. Only at $U_c(T)$ (the blue line in Fig.~\ref{fig:sketch}), both solutions have exactly the same free energy, this is where the phase transition occurs. At $U_{c1}$ ($U_{c2}$) the unstable and insulating (metallic) fixed point merge, so that $\lambda=1$ at this fixed point, but the metallic (insulating) solution, with $\lambda<1$, is the global minimum of the free energy.

Kotliar et al.~\cite{Kotliar02} predicted a compressibility divergence at the critical end point of the doping-driven Mott transition, $dn/d\mu\rightarrow\infty$.
On first sight, our present eigenvalue analysis seems to imply the same, since the BSE diverges.
However, a divergence in the BSE can be canceled by an exact orthogonality~\cite{Kotliar02,Springer19}, and that is indeed what happens at particle-hole symmetry~\cite{Reitner20}. The eigenvector $V \propto \Delta_m-\Delta_i$ is antisymmetric in $\nu$ and therefore does not contribute to the sum in Eq.~\eqref{eq:dndmutrace}~\cite{Chalupa18,Springer19,Reitner20}, so that $dn/d\mu$, shown in Fig.~\ref{fig:obs}, is finite (and small,~\cite{Hafermann14b}) at the critical end point.
This is consistent with the absence of phase separation at particle-hole symmetry~\cite{Eckstein07}.
A non-divergent compressibility combined with a divergence of the BSE is reminiscent of the zero temperature case~\cite{Krien19}, in other words, both critical end points of the particle-hole symmetric Mott transition are characterized by a divergent BSE without a divergence in $dn/d\mu$.

The situation away from particle-hole symmetry is more complicated because of the complex-valuedness of the Green's functions~\cite{Reitner20}. 
The antiferromagnetic transition in DMFT~\cite{Jarrell92} --- which occurs when the assumption of paramagnetism is lifted --- can also be analyzed along the lines of the current work as a divergence of the BSE in the magnetic channel.
An important open question is the generalization of our analysis to
the instabilities found in multi-orbital Hund's physics~\cite{Werner08,Haule09,deMedici09,deMedici11,Werner12,Stadler15,Medici17,Villar18}, and more generally to systems that show phase separation~\cite{Grilli91,Majumdar94,Tandon99,Held01,Kotliar02,Capone04,Eckstein07,Aichhorn07,Lupi10,Otsuki14,Yee15}. 

In conclusion, we identified the non-local Bethe-Salpeter kernel with the Jacobian of the DMFT fixed point equation and with the curvature of the free energy functional. Near the critical end point of the finite temperature correlation-driven Mott transition the BSE diverges. The eigenvector corresponding to the divergence relates the insulating and metallic solutions that exist below the critical temperature. Particle-hole symmetry then implies that this eigenvector is antisymmetric and does not contribute to the compressibility~\cite{Reitner20}, which remains finite.

\acknowledgments

The authors thank M. Capone, P. Chalupa, H. Hafermann, M. Katsnelson, A. Lichtenstein, M. Sch\"uler, A. Toschi, A. Valli and T. Wehling for stimulating discussions.
E.G.C.P.v.L. is supported by the Zentrale Forschungsf\"orderung of the Universit\"at Bremen. F.K. acknowledges financial support from the Slovenian Research Agency under project number N1-0088. A.K. acknowledges partial financial support within the state assignment of Ministry of Science and Higher Education of the Russian Federation (theme “Quant” No. AAAA-A18-118020190095-4) and RFBR grant 20-02-00252.

\bibliography{references}

\begin{thebibliography}{89}%
\makeatletter
\providecommand \@ifxundefined [1]{%
 \@ifx{#1\undefined}
}%
\providecommand \@ifnum [1]{%
 \ifnum #1\expandafter \@firstoftwo
 \else \expandafter \@secondoftwo
 \fi
}%
\providecommand \@ifx [1]{%
 \ifx #1\expandafter \@firstoftwo
 \else \expandafter \@secondoftwo
 \fi
}%
\providecommand \natexlab [1]{#1}%
\providecommand \enquote  [1]{``#1''}%
\providecommand \bibnamefont  [1]{#1}%
\providecommand \bibfnamefont [1]{#1}%
\providecommand \citenamefont [1]{#1}%
\providecommand \href@noop [0]{\@secondoftwo}%
\providecommand \href [0]{\begingroup \@sanitize@url \@href}%
\providecommand \@href[1]{\@@startlink{#1}\@@href}%
\providecommand \@@href[1]{\endgroup#1\@@endlink}%
\providecommand \@sanitize@url [0]{\catcode `\\12\catcode `\$12\catcode
  `\&12\catcode `\#12\catcode `\^12\catcode `\_12\catcode `\%12\relax}%
\providecommand \@@startlink[1]{}%
\providecommand \@@endlink[0]{}%
\providecommand \url  [0]{\begingroup\@sanitize@url \@url }%
\providecommand \@url [1]{\endgroup\@href {#1}{\urlprefix }}%
\providecommand \urlprefix  [0]{URL }%
\providecommand \Eprint [0]{\href }%
\providecommand \doibase [0]{http://dx.doi.org/}%
\providecommand \selectlanguage [0]{\@gobble}%
\providecommand \bibinfo  [0]{\@secondoftwo}%
\providecommand \bibfield  [0]{\@secondoftwo}%
\providecommand \translation [1]{[#1]}%
\providecommand \BibitemOpen [0]{}%
\providecommand \bibitemStop [0]{}%
\providecommand \bibitemNoStop [0]{.\EOS\space}%
\providecommand \EOS [0]{\spacefactor3000\relax}%
\providecommand \BibitemShut  [1]{\csname bibitem#1\endcsname}%
\let\auto@bib@innerbib\@empty
\bibitem [{\citenamefont {Imada}\ \emph {et~al.}(1998)\citenamefont {Imada},
  \citenamefont {Fujimori},\ and\ \citenamefont {Tokura}}]{Imada1998}%
  \BibitemOpen
  \bibfield  {author} {\bibinfo {author} {\bibfnamefont {M.}~\bibnamefont
  {Imada}}, \bibinfo {author} {\bibfnamefont {A.}~\bibnamefont {Fujimori}}, \
  and\ \bibinfo {author} {\bibfnamefont {Y.}~\bibnamefont {Tokura}},\ }\href
  {\doibase 10.1103/RevModPhys.70.1039} {\bibfield  {journal} {\bibinfo
  {journal} {Rev. Mod. Phys.}\ }\textbf {\bibinfo {volume} {70}},\ \bibinfo
  {pages} {1039} (\bibinfo {year} {1998})}\BibitemShut {NoStop}%
\bibitem [{\citenamefont {Hubbard}(1963)}]{Hubbard63}%
  \BibitemOpen
  \bibfield  {author} {\bibinfo {author} {\bibfnamefont {J.}~\bibnamefont
  {Hubbard}},\ }\href {\doibase 10.1098/rspa.1963.0204} {\bibfield  {journal}
  {\bibinfo  {journal} {Proc. R. Soc. A.}\ }\textbf {\bibinfo {volume} {276}},\
  \bibinfo {pages} {238} (\bibinfo {year} {1963})}\BibitemShut {NoStop}%
\bibitem [{\citenamefont {Kanamori}(1963)}]{Kanamori63}%
  \BibitemOpen
  \bibfield  {author} {\bibinfo {author} {\bibfnamefont {J.}~\bibnamefont
  {Kanamori}},\ }\href@noop {} {\bibfield  {journal} {\bibinfo  {journal}
  {Prog. Theor. Phys.}\ }\textbf {\bibinfo {volume} {30}},\ \bibinfo {pages}
  {275} (\bibinfo {year} {1963})}\BibitemShut {NoStop}%
\bibitem [{\citenamefont {Gutzwiller}(1963)}]{Gutzwiller63}%
  \BibitemOpen
  \bibfield  {author} {\bibinfo {author} {\bibfnamefont {M.~C.}\ \bibnamefont
  {Gutzwiller}},\ }\href {\doibase 10.1103/PhysRevLett.10.159} {\bibfield
  {journal} {\bibinfo  {journal} {Phys. Rev. Lett.}\ }\textbf {\bibinfo
  {volume} {10}},\ \bibinfo {pages} {159} (\bibinfo {year} {1963})}\BibitemShut
  {NoStop}%
\bibitem [{\citenamefont {Hubbard}(1964)}]{Hubbard64}%
  \BibitemOpen
  \bibfield  {author} {\bibinfo {author} {\bibfnamefont {J.}~\bibnamefont
  {Hubbard}},\ }\href {\doibase 10.1098/rspa.1964.0190} {\bibfield  {journal}
  {\bibinfo  {journal} {Proc. R. Soc. A.}\ }\textbf {\bibinfo {volume} {281}},\
  \bibinfo {pages} {401} (\bibinfo {year} {1964})}\BibitemShut {NoStop}%
\bibitem [{\citenamefont {J{\"o}rdens}\ \emph {et~al.}(2008)\citenamefont
  {J{\"o}rdens}, \citenamefont {Strohmaier}, \citenamefont {G{\"u}nter},
  \citenamefont {Moritz},\ and\ \citenamefont {Esslinger}}]{Jordens08}%
  \BibitemOpen
  \bibfield  {author} {\bibinfo {author} {\bibfnamefont {R.}~\bibnamefont
  {J{\"o}rdens}}, \bibinfo {author} {\bibfnamefont {N.}~\bibnamefont
  {Strohmaier}}, \bibinfo {author} {\bibfnamefont {K.}~\bibnamefont
  {G{\"u}nter}}, \bibinfo {author} {\bibfnamefont {H.}~\bibnamefont {Moritz}},
  \ and\ \bibinfo {author} {\bibfnamefont {T.}~\bibnamefont {Esslinger}},\
  }\href@noop {} {\bibfield  {journal} {\bibinfo  {journal} {Nature}\ }\textbf
  {\bibinfo {volume} {455}},\ \bibinfo {pages} {204} (\bibinfo {year}
  {2008})}\BibitemShut {NoStop}%
\bibitem [{\citenamefont {Schneider}\ \emph {et~al.}(2008)\citenamefont
  {Schneider}, \citenamefont {Hackerm{\"u}ller}, \citenamefont {Will},
  \citenamefont {Best}, \citenamefont {Bloch}, \citenamefont {Costi},
  \citenamefont {Helmes}, \citenamefont {Rasch},\ and\ \citenamefont
  {Rosch}}]{Schneider08}%
  \BibitemOpen
  \bibfield  {author} {\bibinfo {author} {\bibfnamefont {U.}~\bibnamefont
  {Schneider}}, \bibinfo {author} {\bibfnamefont {L.}~\bibnamefont
  {Hackerm{\"u}ller}}, \bibinfo {author} {\bibfnamefont {S.}~\bibnamefont
  {Will}}, \bibinfo {author} {\bibfnamefont {T.}~\bibnamefont {Best}}, \bibinfo
  {author} {\bibfnamefont {I.}~\bibnamefont {Bloch}}, \bibinfo {author}
  {\bibfnamefont {T.~A.}\ \bibnamefont {Costi}}, \bibinfo {author}
  {\bibfnamefont {R.~W.}\ \bibnamefont {Helmes}}, \bibinfo {author}
  {\bibfnamefont {D.}~\bibnamefont {Rasch}}, \ and\ \bibinfo {author}
  {\bibfnamefont {A.}~\bibnamefont {Rosch}},\ }\href {\doibase
  10.1126/science.1165449} {\bibfield  {journal} {\bibinfo  {journal}
  {Science}\ }\textbf {\bibinfo {volume} {322}},\ \bibinfo {pages} {1520}
  (\bibinfo {year} {2008})}\BibitemShut {NoStop}%
\bibitem [{\citenamefont {J\"ordens}\ \emph {et~al.}(2010)\citenamefont
  {J\"ordens}, \citenamefont {Tarruell}, \citenamefont {Greif}, \citenamefont
  {Uehlinger}, \citenamefont {Strohmaier}, \citenamefont {Moritz},
  \citenamefont {Esslinger}, \citenamefont {De~Leo}, \citenamefont {Kollath},
  \citenamefont {Georges}, \citenamefont {Scarola}, \citenamefont {Pollet},
  \citenamefont {Burovski}, \citenamefont {Kozik},\ and\ \citenamefont
  {Troyer}}]{Jordens10}%
  \BibitemOpen
  \bibfield  {author} {\bibinfo {author} {\bibfnamefont {R.}~\bibnamefont
  {J\"ordens}}, \bibinfo {author} {\bibfnamefont {L.}~\bibnamefont {Tarruell}},
  \bibinfo {author} {\bibfnamefont {D.}~\bibnamefont {Greif}}, \bibinfo
  {author} {\bibfnamefont {T.}~\bibnamefont {Uehlinger}}, \bibinfo {author}
  {\bibfnamefont {N.}~\bibnamefont {Strohmaier}}, \bibinfo {author}
  {\bibfnamefont {H.}~\bibnamefont {Moritz}}, \bibinfo {author} {\bibfnamefont
  {T.}~\bibnamefont {Esslinger}}, \bibinfo {author} {\bibfnamefont
  {L.}~\bibnamefont {De~Leo}}, \bibinfo {author} {\bibfnamefont
  {C.}~\bibnamefont {Kollath}}, \bibinfo {author} {\bibfnamefont
  {A.}~\bibnamefont {Georges}}, \bibinfo {author} {\bibfnamefont
  {V.}~\bibnamefont {Scarola}}, \bibinfo {author} {\bibfnamefont
  {L.}~\bibnamefont {Pollet}}, \bibinfo {author} {\bibfnamefont
  {E.}~\bibnamefont {Burovski}}, \bibinfo {author} {\bibfnamefont
  {E.}~\bibnamefont {Kozik}}, \ and\ \bibinfo {author} {\bibfnamefont
  {M.}~\bibnamefont {Troyer}},\ }\href {\doibase
  10.1103/PhysRevLett.104.180401} {\bibfield  {journal} {\bibinfo  {journal}
  {Phys. Rev. Lett.}\ }\textbf {\bibinfo {volume} {104}},\ \bibinfo {pages}
  {180401} (\bibinfo {year} {2010})}\BibitemShut {NoStop}%
\bibitem [{\citenamefont {Duarte}\ \emph {et~al.}(2015)\citenamefont {Duarte},
  \citenamefont {Hart}, \citenamefont {Yang}, \citenamefont {Liu},
  \citenamefont {Paiva}, \citenamefont {Khatami}, \citenamefont {Scalettar},
  \citenamefont {Trivedi},\ and\ \citenamefont {Hulet}}]{Duarte15}%
  \BibitemOpen
  \bibfield  {author} {\bibinfo {author} {\bibfnamefont {P.~M.}\ \bibnamefont
  {Duarte}}, \bibinfo {author} {\bibfnamefont {R.~A.}\ \bibnamefont {Hart}},
  \bibinfo {author} {\bibfnamefont {T.-L.}\ \bibnamefont {Yang}}, \bibinfo
  {author} {\bibfnamefont {X.}~\bibnamefont {Liu}}, \bibinfo {author}
  {\bibfnamefont {T.}~\bibnamefont {Paiva}}, \bibinfo {author} {\bibfnamefont
  {E.}~\bibnamefont {Khatami}}, \bibinfo {author} {\bibfnamefont {R.~T.}\
  \bibnamefont {Scalettar}}, \bibinfo {author} {\bibfnamefont {N.}~\bibnamefont
  {Trivedi}}, \ and\ \bibinfo {author} {\bibfnamefont {R.~G.}\ \bibnamefont
  {Hulet}},\ }\href {\doibase 10.1103/PhysRevLett.114.070403} {\bibfield
  {journal} {\bibinfo  {journal} {Phys. Rev. Lett.}\ }\textbf {\bibinfo
  {volume} {114}},\ \bibinfo {pages} {070403} (\bibinfo {year}
  {2015})}\BibitemShut {NoStop}%
\bibitem [{\citenamefont {Greif}\ \emph {et~al.}(2016)\citenamefont {Greif},
  \citenamefont {Parsons}, \citenamefont {Mazurenko}, \citenamefont {Chiu},
  \citenamefont {Blatt}, \citenamefont {Huber}, \citenamefont {Ji},\ and\
  \citenamefont {Greiner}}]{Greif16}%
  \BibitemOpen
  \bibfield  {author} {\bibinfo {author} {\bibfnamefont {D.}~\bibnamefont
  {Greif}}, \bibinfo {author} {\bibfnamefont {M.~F.}\ \bibnamefont {Parsons}},
  \bibinfo {author} {\bibfnamefont {A.}~\bibnamefont {Mazurenko}}, \bibinfo
  {author} {\bibfnamefont {C.~S.}\ \bibnamefont {Chiu}}, \bibinfo {author}
  {\bibfnamefont {S.}~\bibnamefont {Blatt}}, \bibinfo {author} {\bibfnamefont
  {F.}~\bibnamefont {Huber}}, \bibinfo {author} {\bibfnamefont
  {G.}~\bibnamefont {Ji}}, \ and\ \bibinfo {author} {\bibfnamefont
  {M.}~\bibnamefont {Greiner}},\ }\href {\doibase 10.1126/science.aad9041}
  {\bibfield  {journal} {\bibinfo  {journal} {Science}\ }\textbf {\bibinfo
  {volume} {351}},\ \bibinfo {pages} {953} (\bibinfo {year}
  {2016})}\BibitemShut {NoStop}%
\bibitem [{\citenamefont {Metzner}\ and\ \citenamefont
  {Vollhardt}(1989)}]{Metzner89}%
  \BibitemOpen
  \bibfield  {author} {\bibinfo {author} {\bibfnamefont {W.}~\bibnamefont
  {Metzner}}\ and\ \bibinfo {author} {\bibfnamefont {D.}~\bibnamefont
  {Vollhardt}},\ }\href {\doibase 10.1103/PhysRevLett.62.324} {\bibfield
  {journal} {\bibinfo  {journal} {Phys. Rev. Lett.}\ }\textbf {\bibinfo
  {volume} {62}},\ \bibinfo {pages} {324} (\bibinfo {year} {1989})}\BibitemShut
  {NoStop}%
\bibitem [{\citenamefont {Georges}\ \emph {et~al.}(1996)\citenamefont
  {Georges}, \citenamefont {Kotliar}, \citenamefont {Krauth},\ and\
  \citenamefont {Rozenberg}}]{Georges96}%
  \BibitemOpen
  \bibfield  {author} {\bibinfo {author} {\bibfnamefont {A.}~\bibnamefont
  {Georges}}, \bibinfo {author} {\bibfnamefont {G.}~\bibnamefont {Kotliar}},
  \bibinfo {author} {\bibfnamefont {W.}~\bibnamefont {Krauth}}, \ and\ \bibinfo
  {author} {\bibfnamefont {M.~J.}\ \bibnamefont {Rozenberg}},\ }\href {\doibase
  10.1103/RevModPhys.68.13} {\bibfield  {journal} {\bibinfo  {journal} {Rev.
  Mod. Phys.}\ }\textbf {\bibinfo {volume} {68}},\ \bibinfo {pages} {13}
  (\bibinfo {year} {1996})}\BibitemShut {NoStop}%
\bibitem [{Note1()}]{Note1}%
  \BibitemOpen
  \bibinfo {note} {New perspectives still appear, such as topological views on
  the transition~\cite {Logan15,Sen20}.}\BibitemShut {Stop}%
\bibitem [{\citenamefont {Jarrell}(1992)}]{Jarrell92}%
  \BibitemOpen
  \bibfield  {author} {\bibinfo {author} {\bibfnamefont {M.}~\bibnamefont
  {Jarrell}},\ }\href {\doibase 10.1103/PhysRevLett.69.168} {\bibfield
  {journal} {\bibinfo  {journal} {Phys. Rev. Lett.}\ }\textbf {\bibinfo
  {volume} {69}},\ \bibinfo {pages} {168} (\bibinfo {year} {1992})}\BibitemShut
  {NoStop}%
\bibitem [{\citenamefont {Georges}\ and\ \citenamefont
  {Krauth}(1992)}]{Georges92}%
  \BibitemOpen
  \bibfield  {author} {\bibinfo {author} {\bibfnamefont {A.}~\bibnamefont
  {Georges}}\ and\ \bibinfo {author} {\bibfnamefont {W.}~\bibnamefont
  {Krauth}},\ }\href {\doibase 10.1103/PhysRevLett.69.1240} {\bibfield
  {journal} {\bibinfo  {journal} {Phys. Rev. Lett.}\ }\textbf {\bibinfo
  {volume} {69}},\ \bibinfo {pages} {1240} (\bibinfo {year}
  {1992})}\BibitemShut {NoStop}%
\bibitem [{\citenamefont {Georges}\ and\ \citenamefont
  {Krauth}(1993)}]{Georges93}%
  \BibitemOpen
  \bibfield  {author} {\bibinfo {author} {\bibfnamefont {A.}~\bibnamefont
  {Georges}}\ and\ \bibinfo {author} {\bibfnamefont {W.}~\bibnamefont
  {Krauth}},\ }\href {\doibase 10.1103/PhysRevB.48.7167} {\bibfield  {journal}
  {\bibinfo  {journal} {Phys. Rev. B}\ }\textbf {\bibinfo {volume} {48}},\
  \bibinfo {pages} {7167} (\bibinfo {year} {1993})}\BibitemShut {NoStop}%
\bibitem [{\citenamefont {Zhang}\ \emph {et~al.}(1993)\citenamefont {Zhang},
  \citenamefont {Rozenberg},\ and\ \citenamefont {Kotliar}}]{Zhang93}%
  \BibitemOpen
  \bibfield  {author} {\bibinfo {author} {\bibfnamefont {X.~Y.}\ \bibnamefont
  {Zhang}}, \bibinfo {author} {\bibfnamefont {M.~J.}\ \bibnamefont
  {Rozenberg}}, \ and\ \bibinfo {author} {\bibfnamefont {G.}~\bibnamefont
  {Kotliar}},\ }\href {\doibase 10.1103/PhysRevLett.70.1666} {\bibfield
  {journal} {\bibinfo  {journal} {Phys. Rev. Lett.}\ }\textbf {\bibinfo
  {volume} {70}},\ \bibinfo {pages} {1666} (\bibinfo {year}
  {1993})}\BibitemShut {NoStop}%
\bibitem [{\citenamefont {Rozenberg}\ \emph {et~al.}(1994)\citenamefont
  {Rozenberg}, \citenamefont {Kotliar},\ and\ \citenamefont
  {Zhang}}]{Rozenberg94}%
  \BibitemOpen
  \bibfield  {author} {\bibinfo {author} {\bibfnamefont {M.~J.}\ \bibnamefont
  {Rozenberg}}, \bibinfo {author} {\bibfnamefont {G.}~\bibnamefont {Kotliar}},
  \ and\ \bibinfo {author} {\bibfnamefont {X.~Y.}\ \bibnamefont {Zhang}},\
  }\href {\doibase 10.1103/PhysRevB.49.10181} {\bibfield  {journal} {\bibinfo
  {journal} {Phys. Rev. B}\ }\textbf {\bibinfo {volume} {49}},\ \bibinfo
  {pages} {10181} (\bibinfo {year} {1994})}\BibitemShut {NoStop}%
\bibitem [{\citenamefont {Noack}\ and\ \citenamefont
  {Gebhard}(1999)}]{Noack99}%
  \BibitemOpen
  \bibfield  {author} {\bibinfo {author} {\bibfnamefont {R.~M.}\ \bibnamefont
  {Noack}}\ and\ \bibinfo {author} {\bibfnamefont {F.}~\bibnamefont
  {Gebhard}},\ }\href {\doibase 10.1103/PhysRevLett.82.1915} {\bibfield
  {journal} {\bibinfo  {journal} {Phys. Rev. Lett.}\ }\textbf {\bibinfo
  {volume} {82}},\ \bibinfo {pages} {1915} (\bibinfo {year}
  {1999})}\BibitemShut {NoStop}%
\bibitem [{\citenamefont {Bulla}(1999)}]{Bulla99}%
  \BibitemOpen
  \bibfield  {author} {\bibinfo {author} {\bibfnamefont {R.}~\bibnamefont
  {Bulla}},\ }\href {\doibase 10.1103/PhysRevLett.83.136} {\bibfield  {journal}
  {\bibinfo  {journal} {Phys. Rev. Lett.}\ }\textbf {\bibinfo {volume} {83}},\
  \bibinfo {pages} {136} (\bibinfo {year} {1999})}\BibitemShut {NoStop}%
\bibitem [{\citenamefont {Bl\"umer}(2002)}]{Blumerphd}%
  \BibitemOpen
  \bibfield  {author} {\bibinfo {author} {\bibfnamefont {N.}~\bibnamefont
  {Bl\"umer}},\ }\emph {\bibinfo {title} {Mott-Hubbard Metal-Insulator
  Transition and Optical Conductivity in High Dimensions}},\ \href@noop {}
  {Ph.D. thesis},\ \bibinfo  {school} {University of Augsburg} (\bibinfo {year}
  {2002})\BibitemShut {NoStop}%
\bibitem [{\citenamefont {Eckstein}\ \emph {et~al.}(2007)\citenamefont
  {Eckstein}, \citenamefont {Kollar}, \citenamefont {Potthoff},\ and\
  \citenamefont {Vollhardt}}]{Eckstein07}%
  \BibitemOpen
  \bibfield  {author} {\bibinfo {author} {\bibfnamefont {M.}~\bibnamefont
  {Eckstein}}, \bibinfo {author} {\bibfnamefont {M.}~\bibnamefont {Kollar}},
  \bibinfo {author} {\bibfnamefont {M.}~\bibnamefont {Potthoff}}, \ and\
  \bibinfo {author} {\bibfnamefont {D.}~\bibnamefont {Vollhardt}},\ }\href
  {\doibase 10.1103/PhysRevB.75.125103} {\bibfield  {journal} {\bibinfo
  {journal} {Phys. Rev. B}\ }\textbf {\bibinfo {volume} {75}},\ \bibinfo
  {pages} {125103} (\bibinfo {year} {2007})}\BibitemShut {NoStop}%
\bibitem [{\citenamefont {Strand}\ \emph {et~al.}(2011)\citenamefont {Strand},
  \citenamefont {Sabashvili}, \citenamefont {Granath}, \citenamefont
  {Hellsing},\ and\ \citenamefont {\"Ostlund}}]{Strand11}%
  \BibitemOpen
  \bibfield  {author} {\bibinfo {author} {\bibfnamefont {H.~U.~R.}\
  \bibnamefont {Strand}}, \bibinfo {author} {\bibfnamefont {A.}~\bibnamefont
  {Sabashvili}}, \bibinfo {author} {\bibfnamefont {M.}~\bibnamefont {Granath}},
  \bibinfo {author} {\bibfnamefont {B.}~\bibnamefont {Hellsing}}, \ and\
  \bibinfo {author} {\bibfnamefont {S.}~\bibnamefont {\"Ostlund}},\ }\href
  {\doibase 10.1103/PhysRevB.83.205136} {\bibfield  {journal} {\bibinfo
  {journal} {Phys. Rev. B}\ }\textbf {\bibinfo {volume} {83}},\ \bibinfo
  {pages} {205136} (\bibinfo {year} {2011})}\BibitemShut {NoStop}%
\bibitem [{\citenamefont {Sch\"afer}\ \emph {et~al.}(2015)\citenamefont
  {Sch\"afer}, \citenamefont {Geles}, \citenamefont {Rost}, \citenamefont
  {Rohringer}, \citenamefont {Arrigoni}, \citenamefont {Held}, \citenamefont
  {Bl\"umer}, \citenamefont {Aichhorn},\ and\ \citenamefont
  {Toschi}}]{Schafer15}%
  \BibitemOpen
  \bibfield  {author} {\bibinfo {author} {\bibfnamefont {T.}~\bibnamefont
  {Sch\"afer}}, \bibinfo {author} {\bibfnamefont {F.}~\bibnamefont {Geles}},
  \bibinfo {author} {\bibfnamefont {D.}~\bibnamefont {Rost}}, \bibinfo {author}
  {\bibfnamefont {G.}~\bibnamefont {Rohringer}}, \bibinfo {author}
  {\bibfnamefont {E.}~\bibnamefont {Arrigoni}}, \bibinfo {author}
  {\bibfnamefont {K.}~\bibnamefont {Held}}, \bibinfo {author} {\bibfnamefont
  {N.}~\bibnamefont {Bl\"umer}}, \bibinfo {author} {\bibfnamefont
  {M.}~\bibnamefont {Aichhorn}}, \ and\ \bibinfo {author} {\bibfnamefont
  {A.}~\bibnamefont {Toschi}},\ }\href {\doibase 10.1103/PhysRevB.91.125109}
  {\bibfield  {journal} {\bibinfo  {journal} {Phys. Rev. B}\ }\textbf {\bibinfo
  {volume} {91}},\ \bibinfo {pages} {125109} (\bibinfo {year}
  {2015})}\BibitemShut {NoStop}%
\bibitem [{\citenamefont {Khurana}(1990)}]{Khurana90}%
  \BibitemOpen
  \bibfield  {author} {\bibinfo {author} {\bibfnamefont {A.}~\bibnamefont
  {Khurana}},\ }\href {\doibase 10.1103/PhysRevLett.64.1990} {\bibfield
  {journal} {\bibinfo  {journal} {Phys. Rev. Lett.}\ }\textbf {\bibinfo
  {volume} {64}},\ \bibinfo {pages} {1990} (\bibinfo {year}
  {1990})}\BibitemShut {NoStop}%
\bibitem [{\citenamefont {Zlatic}\ and\ \citenamefont
  {Horvatic}(1990)}]{Zlatic90}%
  \BibitemOpen
  \bibfield  {author} {\bibinfo {author} {\bibfnamefont {V.}~\bibnamefont
  {Zlatic}}\ and\ \bibinfo {author} {\bibfnamefont {B.}~\bibnamefont
  {Horvatic}},\ }\href {\doibase https://doi.org/10.1016/0038-1098(90)90282-G}
  {\bibfield  {journal} {\bibinfo  {journal} {Solid State Communications}\
  }\textbf {\bibinfo {volume} {75}},\ \bibinfo {pages} {263 } (\bibinfo {year}
  {1990})}\BibitemShut {NoStop}%
\bibitem [{\citenamefont {Pruschke}\ \emph {et~al.}(1993)\citenamefont
  {Pruschke}, \citenamefont {Cox},\ and\ \citenamefont {Jarrell}}]{Pruschke93}%
  \BibitemOpen
  \bibfield  {author} {\bibinfo {author} {\bibfnamefont {T.}~\bibnamefont
  {Pruschke}}, \bibinfo {author} {\bibfnamefont {D.~L.}\ \bibnamefont {Cox}}, \
  and\ \bibinfo {author} {\bibfnamefont {M.}~\bibnamefont {Jarrell}},\ }\href
  {\doibase 10.1103/PhysRevB.47.3553} {\bibfield  {journal} {\bibinfo
  {journal} {Phys. Rev. B}\ }\textbf {\bibinfo {volume} {47}},\ \bibinfo
  {pages} {3553} (\bibinfo {year} {1993})}\BibitemShut {NoStop}%
\bibitem [{\citenamefont {Rozenberg}\ \emph {et~al.}(1995)\citenamefont
  {Rozenberg}, \citenamefont {Kotliar}, \citenamefont {Kajueter}, \citenamefont
  {Thomas}, \citenamefont {Rapkine}, \citenamefont {Honig},\ and\ \citenamefont
  {Metcalf}}]{Rozenberg95}%
  \BibitemOpen
  \bibfield  {author} {\bibinfo {author} {\bibfnamefont {M.~J.}\ \bibnamefont
  {Rozenberg}}, \bibinfo {author} {\bibfnamefont {G.}~\bibnamefont {Kotliar}},
  \bibinfo {author} {\bibfnamefont {H.}~\bibnamefont {Kajueter}}, \bibinfo
  {author} {\bibfnamefont {G.~A.}\ \bibnamefont {Thomas}}, \bibinfo {author}
  {\bibfnamefont {D.~H.}\ \bibnamefont {Rapkine}}, \bibinfo {author}
  {\bibfnamefont {J.~M.}\ \bibnamefont {Honig}}, \ and\ \bibinfo {author}
  {\bibfnamefont {P.}~\bibnamefont {Metcalf}},\ }\href {\doibase
  10.1103/PhysRevLett.75.105} {\bibfield  {journal} {\bibinfo  {journal} {Phys.
  Rev. Lett.}\ }\textbf {\bibinfo {volume} {75}},\ \bibinfo {pages} {105}
  (\bibinfo {year} {1995})}\BibitemShut {NoStop}%
\bibitem [{\citenamefont {Brener}\ \emph {et~al.}(2008)\citenamefont {Brener},
  \citenamefont {Hafermann}, \citenamefont {Rubtsov}, \citenamefont
  {Katsnelson},\ and\ \citenamefont {Lichtenstein}}]{Brener08}%
  \BibitemOpen
  \bibfield  {author} {\bibinfo {author} {\bibfnamefont {S.}~\bibnamefont
  {Brener}}, \bibinfo {author} {\bibfnamefont {H.}~\bibnamefont {Hafermann}},
  \bibinfo {author} {\bibfnamefont {A.~N.}\ \bibnamefont {Rubtsov}}, \bibinfo
  {author} {\bibfnamefont {M.~I.}\ \bibnamefont {Katsnelson}}, \ and\ \bibinfo
  {author} {\bibfnamefont {A.~I.}\ \bibnamefont {Lichtenstein}},\ }\href
  {\doibase 10.1103/PhysRevB.77.195105} {\bibfield  {journal} {\bibinfo
  {journal} {Phys. Rev. B}\ }\textbf {\bibinfo {volume} {77}},\ \bibinfo
  {pages} {195105} (\bibinfo {year} {2008})}\BibitemShut {NoStop}%
\bibitem [{\citenamefont {Rohringer}\ \emph {et~al.}(2012)\citenamefont
  {Rohringer}, \citenamefont {Valli},\ and\ \citenamefont
  {Toschi}}]{Rohringer12}%
  \BibitemOpen
  \bibfield  {author} {\bibinfo {author} {\bibfnamefont {G.}~\bibnamefont
  {Rohringer}}, \bibinfo {author} {\bibfnamefont {A.}~\bibnamefont {Valli}}, \
  and\ \bibinfo {author} {\bibfnamefont {A.}~\bibnamefont {Toschi}},\ }\href
  {\doibase 10.1103/PhysRevB.86.125114} {\bibfield  {journal} {\bibinfo
  {journal} {Phys. Rev. B}\ }\textbf {\bibinfo {volume} {86}},\ \bibinfo
  {pages} {125114} (\bibinfo {year} {2012})}\BibitemShut {NoStop}%
\bibitem [{\citenamefont {Boehnke}\ and\ \citenamefont
  {Lechermann}(2012)}]{Boehnke12}%
  \BibitemOpen
  \bibfield  {author} {\bibinfo {author} {\bibfnamefont {L.}~\bibnamefont
  {Boehnke}}\ and\ \bibinfo {author} {\bibfnamefont {F.}~\bibnamefont
  {Lechermann}},\ }\href {\doibase 10.1103/PhysRevB.85.115128} {\bibfield
  {journal} {\bibinfo  {journal} {Phys. Rev. B}\ }\textbf {\bibinfo {volume}
  {85}},\ \bibinfo {pages} {115128} (\bibinfo {year} {2012})}\BibitemShut
  {NoStop}%
\bibitem [{\citenamefont {van Loon}\ \emph {et~al.}(2014)\citenamefont {van
  Loon}, \citenamefont {Hafermann}, \citenamefont {Lichtenstein}, \citenamefont
  {Rubtsov},\ and\ \citenamefont {Katsnelson}}]{vanLoon14b}%
  \BibitemOpen
  \bibfield  {author} {\bibinfo {author} {\bibfnamefont {E.~G. C.~P.}\
  \bibnamefont {van Loon}}, \bibinfo {author} {\bibfnamefont {H.}~\bibnamefont
  {Hafermann}}, \bibinfo {author} {\bibfnamefont {A.~I.}\ \bibnamefont
  {Lichtenstein}}, \bibinfo {author} {\bibfnamefont {A.~N.}\ \bibnamefont
  {Rubtsov}}, \ and\ \bibinfo {author} {\bibfnamefont {M.~I.}\ \bibnamefont
  {Katsnelson}},\ }\href {\doibase 10.1103/PhysRevLett.113.246407} {\bibfield
  {journal} {\bibinfo  {journal} {Phys. Rev. Lett.}\ }\textbf {\bibinfo
  {volume} {113}},\ \bibinfo {pages} {246407} (\bibinfo {year}
  {2014})}\BibitemShut {NoStop}%
\bibitem [{\citenamefont {Geffroy}\ \emph {et~al.}(2019)\citenamefont
  {Geffroy}, \citenamefont {Kaufmann}, \citenamefont {Hariki}, \citenamefont
  {Gunacker}, \citenamefont {Hausoel},\ and\ \citenamefont
  {Kune\ifmmode~\check{s}\else \v{s}\fi{}}}]{Geffroy19}%
  \BibitemOpen
  \bibfield  {author} {\bibinfo {author} {\bibfnamefont {D.}~\bibnamefont
  {Geffroy}}, \bibinfo {author} {\bibfnamefont {J.}~\bibnamefont {Kaufmann}},
  \bibinfo {author} {\bibfnamefont {A.}~\bibnamefont {Hariki}}, \bibinfo
  {author} {\bibfnamefont {P.}~\bibnamefont {Gunacker}}, \bibinfo {author}
  {\bibfnamefont {A.}~\bibnamefont {Hausoel}}, \ and\ \bibinfo {author}
  {\bibfnamefont {J.}~\bibnamefont {Kune\ifmmode~\check{s}\else \v{s}\fi{}}},\
  }\href {\doibase 10.1103/PhysRevLett.122.127601} {\bibfield  {journal}
  {\bibinfo  {journal} {Phys. Rev. Lett.}\ }\textbf {\bibinfo {volume} {122}},\
  \bibinfo {pages} {127601} (\bibinfo {year} {2019})}\BibitemShut {NoStop}%
\bibitem [{\citenamefont {Strand}\ \emph {et~al.}(2019)\citenamefont {Strand},
  \citenamefont {Zingl}, \citenamefont {Wentzell}, \citenamefont {Parcollet},\
  and\ \citenamefont {Georges}}]{Strand19}%
  \BibitemOpen
  \bibfield  {author} {\bibinfo {author} {\bibfnamefont {H.~U.~R.}\
  \bibnamefont {Strand}}, \bibinfo {author} {\bibfnamefont {M.}~\bibnamefont
  {Zingl}}, \bibinfo {author} {\bibfnamefont {N.}~\bibnamefont {Wentzell}},
  \bibinfo {author} {\bibfnamefont {O.}~\bibnamefont {Parcollet}}, \ and\
  \bibinfo {author} {\bibfnamefont {A.}~\bibnamefont {Georges}},\ }\href
  {\doibase 10.1103/PhysRevB.100.125120} {\bibfield  {journal} {\bibinfo
  {journal} {Phys. Rev. B}\ }\textbf {\bibinfo {volume} {100}},\ \bibinfo
  {pages} {125120} (\bibinfo {year} {2019})}\BibitemShut {NoStop}%
\bibitem [{\citenamefont {Krien}\ \emph {et~al.}(2019)\citenamefont {Krien},
  \citenamefont {van Loon}, \citenamefont {Katsnelson}, \citenamefont
  {Lichtenstein},\ and\ \citenamefont {Capone}}]{Krien19}%
  \BibitemOpen
  \bibfield  {author} {\bibinfo {author} {\bibfnamefont {F.}~\bibnamefont
  {Krien}}, \bibinfo {author} {\bibfnamefont {E.~G. C.~P.}\ \bibnamefont {van
  Loon}}, \bibinfo {author} {\bibfnamefont {M.~I.}\ \bibnamefont {Katsnelson}},
  \bibinfo {author} {\bibfnamefont {A.~I.}\ \bibnamefont {Lichtenstein}}, \
  and\ \bibinfo {author} {\bibfnamefont {M.}~\bibnamefont {Capone}},\ }\href
  {\doibase 10.1103/PhysRevB.99.245128} {\bibfield  {journal} {\bibinfo
  {journal} {Phys. Rev. B}\ }\textbf {\bibinfo {volume} {99}},\ \bibinfo
  {pages} {245128} (\bibinfo {year} {2019})}\BibitemShut {NoStop}%
\bibitem [{\citenamefont {Melnick}\ and\ \citenamefont
  {Kotliar}(2020)}]{Melnick20}%
  \BibitemOpen
  \bibfield  {author} {\bibinfo {author} {\bibfnamefont {C.}~\bibnamefont
  {Melnick}}\ and\ \bibinfo {author} {\bibfnamefont {G.}~\bibnamefont
  {Kotliar}},\ }\href {\doibase 10.1103/PhysRevB.101.165105} {\bibfield
  {journal} {\bibinfo  {journal} {Phys. Rev. B}\ }\textbf {\bibinfo {volume}
  {101}},\ \bibinfo {pages} {165105} (\bibinfo {year} {2020})}\BibitemShut
  {NoStop}%
\bibitem [{\citenamefont {Rubtsov}\ \emph {et~al.}(2005)\citenamefont
  {Rubtsov}, \citenamefont {Savkin},\ and\ \citenamefont
  {Lichtenstein}}]{Rubtsov05}%
  \BibitemOpen
  \bibfield  {author} {\bibinfo {author} {\bibfnamefont {A.~N.}\ \bibnamefont
  {Rubtsov}}, \bibinfo {author} {\bibfnamefont {V.~V.}\ \bibnamefont {Savkin}},
  \ and\ \bibinfo {author} {\bibfnamefont {A.~I.}\ \bibnamefont
  {Lichtenstein}},\ }\href {\doibase 10.1103/PhysRevB.72.035122} {\bibfield
  {journal} {\bibinfo  {journal} {Phys. Rev. B}\ }\textbf {\bibinfo {volume}
  {72}},\ \bibinfo {pages} {035122} (\bibinfo {year} {2005})}\BibitemShut
  {NoStop}%
\bibitem [{\citenamefont {Werner}\ \emph {et~al.}(2006)\citenamefont {Werner},
  \citenamefont {Comanac}, \citenamefont {de' Medici}, \citenamefont {Troyer},\
  and\ \citenamefont {Millis}}]{Werner06}%
  \BibitemOpen
  \bibfield  {author} {\bibinfo {author} {\bibfnamefont {P.}~\bibnamefont
  {Werner}}, \bibinfo {author} {\bibfnamefont {A.}~\bibnamefont {Comanac}},
  \bibinfo {author} {\bibfnamefont {L.}~\bibnamefont {de' Medici}}, \bibinfo
  {author} {\bibfnamefont {M.}~\bibnamefont {Troyer}}, \ and\ \bibinfo {author}
  {\bibfnamefont {A.~J.}\ \bibnamefont {Millis}},\ }\href {\doibase
  10.1103/PhysRevLett.97.076405} {\bibfield  {journal} {\bibinfo  {journal}
  {Phys. Rev. Lett.}\ }\textbf {\bibinfo {volume} {97}},\ \bibinfo {pages}
  {076405} (\bibinfo {year} {2006})}\BibitemShut {NoStop}%
\bibitem [{\citenamefont {Gull}\ \emph {et~al.}(2011)\citenamefont {Gull},
  \citenamefont {Millis}, \citenamefont {Lichtenstein}, \citenamefont
  {Rubtsov}, \citenamefont {Troyer},\ and\ \citenamefont {Werner}}]{Gull11}%
  \BibitemOpen
  \bibfield  {author} {\bibinfo {author} {\bibfnamefont {E.}~\bibnamefont
  {Gull}}, \bibinfo {author} {\bibfnamefont {A.~J.}\ \bibnamefont {Millis}},
  \bibinfo {author} {\bibfnamefont {A.~I.}\ \bibnamefont {Lichtenstein}},
  \bibinfo {author} {\bibfnamefont {A.~N.}\ \bibnamefont {Rubtsov}}, \bibinfo
  {author} {\bibfnamefont {M.}~\bibnamefont {Troyer}}, \ and\ \bibinfo {author}
  {\bibfnamefont {P.}~\bibnamefont {Werner}},\ }\href {\doibase
  10.1103/RevModPhys.83.349} {\bibfield  {journal} {\bibinfo  {journal} {Rev.
  Mod. Phys.}\ }\textbf {\bibinfo {volume} {83}},\ \bibinfo {pages} {349}
  (\bibinfo {year} {2011})}\BibitemShut {NoStop}%
\bibitem [{\citenamefont {Sch\"afer}\ \emph {et~al.}(2013)\citenamefont
  {Sch\"afer}, \citenamefont {Rohringer}, \citenamefont {Gunnarsson},
  \citenamefont {Ciuchi}, \citenamefont {Sangiovanni},\ and\ \citenamefont
  {Toschi}}]{Schafer13}%
  \BibitemOpen
  \bibfield  {author} {\bibinfo {author} {\bibfnamefont {T.}~\bibnamefont
  {Sch\"afer}}, \bibinfo {author} {\bibfnamefont {G.}~\bibnamefont
  {Rohringer}}, \bibinfo {author} {\bibfnamefont {O.}~\bibnamefont
  {Gunnarsson}}, \bibinfo {author} {\bibfnamefont {S.}~\bibnamefont {Ciuchi}},
  \bibinfo {author} {\bibfnamefont {G.}~\bibnamefont {Sangiovanni}}, \ and\
  \bibinfo {author} {\bibfnamefont {A.}~\bibnamefont {Toschi}},\ }\href
  {\doibase 10.1103/PhysRevLett.110.246405} {\bibfield  {journal} {\bibinfo
  {journal} {Phys. Rev. Lett.}\ }\textbf {\bibinfo {volume} {110}},\ \bibinfo
  {pages} {246405} (\bibinfo {year} {2013})}\BibitemShut {NoStop}%
\bibitem [{\citenamefont {Kozik}\ \emph {et~al.}(2015)\citenamefont {Kozik},
  \citenamefont {Ferrero},\ and\ \citenamefont {Georges}}]{Kozik15}%
  \BibitemOpen
  \bibfield  {author} {\bibinfo {author} {\bibfnamefont {E.}~\bibnamefont
  {Kozik}}, \bibinfo {author} {\bibfnamefont {M.}~\bibnamefont {Ferrero}}, \
  and\ \bibinfo {author} {\bibfnamefont {A.}~\bibnamefont {Georges}},\ }\href
  {\doibase 10.1103/PhysRevLett.114.156402} {\bibfield  {journal} {\bibinfo
  {journal} {Phys. Rev. Lett.}\ }\textbf {\bibinfo {volume} {114}},\ \bibinfo
  {pages} {156402} (\bibinfo {year} {2015})}\BibitemShut {NoStop}%
\bibitem [{\citenamefont {Sch\"afer}\ \emph {et~al.}(2016)\citenamefont
  {Sch\"afer}, \citenamefont {Ciuchi}, \citenamefont {Wallerberger},
  \citenamefont {Thunstr\"om}, \citenamefont {Gunnarsson}, \citenamefont
  {Sangiovanni}, \citenamefont {Rohringer},\ and\ \citenamefont
  {Toschi}}]{Schafer16}%
  \BibitemOpen
  \bibfield  {author} {\bibinfo {author} {\bibfnamefont {T.}~\bibnamefont
  {Sch\"afer}}, \bibinfo {author} {\bibfnamefont {S.}~\bibnamefont {Ciuchi}},
  \bibinfo {author} {\bibfnamefont {M.}~\bibnamefont {Wallerberger}}, \bibinfo
  {author} {\bibfnamefont {P.}~\bibnamefont {Thunstr\"om}}, \bibinfo {author}
  {\bibfnamefont {O.}~\bibnamefont {Gunnarsson}}, \bibinfo {author}
  {\bibfnamefont {G.}~\bibnamefont {Sangiovanni}}, \bibinfo {author}
  {\bibfnamefont {G.}~\bibnamefont {Rohringer}}, \ and\ \bibinfo {author}
  {\bibfnamefont {A.}~\bibnamefont {Toschi}},\ }\href {\doibase
  10.1103/PhysRevB.94.235108} {\bibfield  {journal} {\bibinfo  {journal} {Phys.
  Rev. B}\ }\textbf {\bibinfo {volume} {94}},\ \bibinfo {pages} {235108}
  (\bibinfo {year} {2016})}\BibitemShut {NoStop}%
\bibitem [{\citenamefont {Gunnarsson}\ \emph {et~al.}(2017)\citenamefont
  {Gunnarsson}, \citenamefont {Rohringer}, \citenamefont {Sch\"afer},
  \citenamefont {Sangiovanni},\ and\ \citenamefont {Toschi}}]{Gunnarsson17}%
  \BibitemOpen
  \bibfield  {author} {\bibinfo {author} {\bibfnamefont {O.}~\bibnamefont
  {Gunnarsson}}, \bibinfo {author} {\bibfnamefont {G.}~\bibnamefont
  {Rohringer}}, \bibinfo {author} {\bibfnamefont {T.}~\bibnamefont
  {Sch\"afer}}, \bibinfo {author} {\bibfnamefont {G.}~\bibnamefont
  {Sangiovanni}}, \ and\ \bibinfo {author} {\bibfnamefont {A.}~\bibnamefont
  {Toschi}},\ }\href {\doibase 10.1103/PhysRevLett.119.056402} {\bibfield
  {journal} {\bibinfo  {journal} {Phys. Rev. Lett.}\ }\textbf {\bibinfo
  {volume} {119}},\ \bibinfo {pages} {056402} (\bibinfo {year}
  {2017})}\BibitemShut {NoStop}%
\bibitem [{\citenamefont {Chalupa}\ \emph {et~al.}(2020)\citenamefont
  {Chalupa}, \citenamefont {Schäfer}, \citenamefont {Reitner}, \citenamefont
  {Springer}, \citenamefont {Andergassen},\ and\ \citenamefont
  {Toschi}}]{Chalupa20}%
  \BibitemOpen
  \bibfield  {author} {\bibinfo {author} {\bibfnamefont {P.}~\bibnamefont
  {Chalupa}}, \bibinfo {author} {\bibfnamefont {T.}~\bibnamefont {Schäfer}},
  \bibinfo {author} {\bibfnamefont {M.}~\bibnamefont {Reitner}}, \bibinfo
  {author} {\bibfnamefont {D.}~\bibnamefont {Springer}}, \bibinfo {author}
  {\bibfnamefont {S.}~\bibnamefont {Andergassen}}, \ and\ \bibinfo {author}
  {\bibfnamefont {A.}~\bibnamefont {Toschi}},\ }\href@noop {} {\enquote
  {\bibinfo {title} {Fingerprints of the local moment formation and its kondo
  screening in the generalized susceptibilities of many-electron problems},}\ }
  (\bibinfo {year} {2020}),\ \Eprint {http://arxiv.org/abs/2003.07829}
  {arXiv:2003.07829 [cond-mat.str-el]} \BibitemShut {NoStop}%
\bibitem [{\citenamefont {Stan}\ \emph {et~al.}(2015)\citenamefont {Stan},
  \citenamefont {Romaniello}, \citenamefont {Rigamonti}, \citenamefont
  {Reining},\ and\ \citenamefont {Berger}}]{Stan15}%
  \BibitemOpen
  \bibfield  {author} {\bibinfo {author} {\bibfnamefont {A.}~\bibnamefont
  {Stan}}, \bibinfo {author} {\bibfnamefont {P.}~\bibnamefont {Romaniello}},
  \bibinfo {author} {\bibfnamefont {S.}~\bibnamefont {Rigamonti}}, \bibinfo
  {author} {\bibfnamefont {L.}~\bibnamefont {Reining}}, \ and\ \bibinfo
  {author} {\bibfnamefont {J.~A.}\ \bibnamefont {Berger}},\ }\href {\doibase
  10.1088/1367-2630/17/9/093045} {\bibfield  {journal} {\bibinfo  {journal}
  {New Journal of Physics}\ }\textbf {\bibinfo {volume} {17}},\ \bibinfo
  {pages} {093045} (\bibinfo {year} {2015})}\BibitemShut {NoStop}%
\bibitem [{\citenamefont {Rossi}\ and\ \citenamefont {Werner}(2015)}]{Rossi15}%
  \BibitemOpen
  \bibfield  {author} {\bibinfo {author} {\bibfnamefont {R.}~\bibnamefont
  {Rossi}}\ and\ \bibinfo {author} {\bibfnamefont {F.}~\bibnamefont {Werner}},\
  }\href {\doibase 10.1088/1751-8113/48/48/485202} {\bibfield  {journal}
  {\bibinfo  {journal} {Journal of Physics A: Mathematical and Theoretical}\
  }\textbf {\bibinfo {volume} {48}},\ \bibinfo {pages} {485202} (\bibinfo
  {year} {2015})}\BibitemShut {NoStop}%
\bibitem [{\citenamefont {Thunstr\"om}\ \emph {et~al.}(2018)\citenamefont
  {Thunstr\"om}, \citenamefont {Gunnarsson}, \citenamefont {Ciuchi},\ and\
  \citenamefont {Rohringer}}]{Thunstrom18}%
  \BibitemOpen
  \bibfield  {author} {\bibinfo {author} {\bibfnamefont {P.}~\bibnamefont
  {Thunstr\"om}}, \bibinfo {author} {\bibfnamefont {O.}~\bibnamefont
  {Gunnarsson}}, \bibinfo {author} {\bibfnamefont {S.}~\bibnamefont {Ciuchi}},
  \ and\ \bibinfo {author} {\bibfnamefont {G.}~\bibnamefont {Rohringer}},\
  }\href {\doibase 10.1103/PhysRevB.98.235107} {\bibfield  {journal} {\bibinfo
  {journal} {Phys. Rev. B}\ }\textbf {\bibinfo {volume} {98}},\ \bibinfo
  {pages} {235107} (\bibinfo {year} {2018})}\BibitemShut {NoStop}%
\bibitem [{\citenamefont {Vu\ifmmode \check{c}\else \v{c}\fi{}i\ifmmode
  \check{c}\else \v{c}\fi{}evi\ifmmode~\acute{c}\else \'{c}\fi{}}\ \emph
  {et~al.}(2018)\citenamefont {Vu\ifmmode \check{c}\else \v{c}\fi{}i\ifmmode
  \check{c}\else \v{c}\fi{}evi\ifmmode~\acute{c}\else \'{c}\fi{}},
  \citenamefont {Wentzell}, \citenamefont {Ferrero},\ and\ \citenamefont
  {Parcollet}}]{Vucicevic18}%
  \BibitemOpen
  \bibfield  {author} {\bibinfo {author} {\bibfnamefont {J.}~\bibnamefont
  {Vu\ifmmode \check{c}\else \v{c}\fi{}i\ifmmode \check{c}\else
  \v{c}\fi{}evi\ifmmode~\acute{c}\else \'{c}\fi{}}}, \bibinfo {author}
  {\bibfnamefont {N.}~\bibnamefont {Wentzell}}, \bibinfo {author}
  {\bibfnamefont {M.}~\bibnamefont {Ferrero}}, \ and\ \bibinfo {author}
  {\bibfnamefont {O.}~\bibnamefont {Parcollet}},\ }\href {\doibase
  10.1103/PhysRevB.97.125141} {\bibfield  {journal} {\bibinfo  {journal} {Phys.
  Rev. B}\ }\textbf {\bibinfo {volume} {97}},\ \bibinfo {pages} {125141}
  (\bibinfo {year} {2018})}\BibitemShut {NoStop}%
\bibitem [{\citenamefont {Tarantino}\ \emph {et~al.}(2017)\citenamefont
  {Tarantino}, \citenamefont {Romaniello}, \citenamefont {Berger},\ and\
  \citenamefont {Reining}}]{Tarantino17}%
  \BibitemOpen
  \bibfield  {author} {\bibinfo {author} {\bibfnamefont {W.}~\bibnamefont
  {Tarantino}}, \bibinfo {author} {\bibfnamefont {P.}~\bibnamefont
  {Romaniello}}, \bibinfo {author} {\bibfnamefont {J.~A.}\ \bibnamefont
  {Berger}}, \ and\ \bibinfo {author} {\bibfnamefont {L.}~\bibnamefont
  {Reining}},\ }\href {\doibase 10.1103/PhysRevB.96.045124} {\bibfield
  {journal} {\bibinfo  {journal} {Phys. Rev. B}\ }\textbf {\bibinfo {volume}
  {96}},\ \bibinfo {pages} {045124} (\bibinfo {year} {2017})}\BibitemShut
  {NoStop}%
\bibitem [{\citenamefont {Chalupa}\ \emph {et~al.}(2018)\citenamefont
  {Chalupa}, \citenamefont {Gunacker}, \citenamefont {Sch\"afer}, \citenamefont
  {Held},\ and\ \citenamefont {Toschi}}]{Chalupa18}%
  \BibitemOpen
  \bibfield  {author} {\bibinfo {author} {\bibfnamefont {P.}~\bibnamefont
  {Chalupa}}, \bibinfo {author} {\bibfnamefont {P.}~\bibnamefont {Gunacker}},
  \bibinfo {author} {\bibfnamefont {T.}~\bibnamefont {Sch\"afer}}, \bibinfo
  {author} {\bibfnamefont {K.}~\bibnamefont {Held}}, \ and\ \bibinfo {author}
  {\bibfnamefont {A.}~\bibnamefont {Toschi}},\ }\href {\doibase
  10.1103/PhysRevB.97.245136} {\bibfield  {journal} {\bibinfo  {journal} {Phys.
  Rev. B}\ }\textbf {\bibinfo {volume} {97}},\ \bibinfo {pages} {245136}
  (\bibinfo {year} {2018})}\BibitemShut {NoStop}%
\bibitem [{\citenamefont {Springer}\ \emph {et~al.}(2020)\citenamefont
  {Springer}, \citenamefont {Chalupa}, \citenamefont {Ciuchi}, \citenamefont
  {Sangiovanni},\ and\ \citenamefont {Toschi}}]{Springer19}%
  \BibitemOpen
  \bibfield  {author} {\bibinfo {author} {\bibfnamefont {D.}~\bibnamefont
  {Springer}}, \bibinfo {author} {\bibfnamefont {P.}~\bibnamefont {Chalupa}},
  \bibinfo {author} {\bibfnamefont {S.}~\bibnamefont {Ciuchi}}, \bibinfo
  {author} {\bibfnamefont {G.}~\bibnamefont {Sangiovanni}}, \ and\ \bibinfo
  {author} {\bibfnamefont {A.}~\bibnamefont {Toschi}},\ }\href {\doibase
  10.1103/PhysRevB.101.155148} {\bibfield  {journal} {\bibinfo  {journal}
  {Phys. Rev. B}\ }\textbf {\bibinfo {volume} {101}},\ \bibinfo {pages}
  {155148} (\bibinfo {year} {2020})}\BibitemShut {NoStop}%
\bibitem [{\citenamefont {Chitra}\ and\ \citenamefont
  {Kotliar}(2001)}]{Chitra01}%
  \BibitemOpen
  \bibfield  {author} {\bibinfo {author} {\bibfnamefont {R.}~\bibnamefont
  {Chitra}}\ and\ \bibinfo {author} {\bibfnamefont {G.}~\bibnamefont
  {Kotliar}},\ }\href {\doibase 10.1103/PhysRevB.63.115110} {\bibfield
  {journal} {\bibinfo  {journal} {Phys. Rev. B}\ }\textbf {\bibinfo {volume}
  {63}},\ \bibinfo {pages} {115110} (\bibinfo {year} {2001})}\BibitemShut
  {NoStop}%
\bibitem [{\citenamefont {Potthoff}(2003)}]{Potthoff03}%
  \BibitemOpen
  \bibfield  {author} {\bibinfo {author} {\bibfnamefont {M.}~\bibnamefont
  {Potthoff}},\ }\href {\doibase 10.1140/epjb/e2003-00121-8} {\bibfield
  {journal} {\bibinfo  {journal} {The European Physical Journal B - Condensed
  Matter and Complex Systems}\ }\textbf {\bibinfo {volume} {32}},\ \bibinfo
  {pages} {429} (\bibinfo {year} {2003})}\BibitemShut {NoStop}%
\bibitem [{\citenamefont {\ifmmode~\check{Z}\else
  \v{Z}\fi{}itko}(2009)}]{Zitko09}%
  \BibitemOpen
  \bibfield  {author} {\bibinfo {author} {\bibfnamefont {R.}~\bibnamefont
  {\ifmmode~\check{Z}\else \v{Z}\fi{}itko}},\ }\href {\doibase
  10.1103/PhysRevB.80.125125} {\bibfield  {journal} {\bibinfo  {journal} {Phys.
  Rev. B}\ }\textbf {\bibinfo {volume} {80}},\ \bibinfo {pages} {125125}
  (\bibinfo {year} {2009})}\BibitemShut {NoStop}%
\bibitem [{\citenamefont {{Reitner}}\ \emph {et~al.}(2020)\citenamefont
  {{Reitner}}, \citenamefont {{Chalupa}}, \citenamefont {{Del Re}},
  \citenamefont {{Springer}}, \citenamefont {{Ciuchi}}, \citenamefont
  {{Sangiovanni}},\ and\ \citenamefont {{Toschi}}}]{Reitner20}%
  \BibitemOpen
  \bibfield  {author} {\bibinfo {author} {\bibfnamefont {M.}~\bibnamefont
  {{Reitner}}}, \bibinfo {author} {\bibfnamefont {P.}~\bibnamefont
  {{Chalupa}}}, \bibinfo {author} {\bibfnamefont {L.}~\bibnamefont {{Del Re}}},
  \bibinfo {author} {\bibfnamefont {D.}~\bibnamefont {{Springer}}}, \bibinfo
  {author} {\bibfnamefont {S.}~\bibnamefont {{Ciuchi}}}, \bibinfo {author}
  {\bibfnamefont {G.}~\bibnamefont {{Sangiovanni}}}, \ and\ \bibinfo {author}
  {\bibfnamefont {A.}~\bibnamefont {{Toschi}}},\ }\href@noop {} {\bibfield
  {journal} {\bibinfo  {journal} {arXiv e-prints}\ ,\ \bibinfo {eid}
  {arXiv:2002.12869}} (\bibinfo {year} {2020})},\ \Eprint
  {http://arxiv.org/abs/2002.12869} {arXiv:2002.12869 [cond-mat.str-el]}
  \BibitemShut {NoStop}%
\bibitem [{\citenamefont {Bauer}\ \emph {et~al.}(2011)\citenamefont {Bauer},
  \citenamefont {Carr}, \citenamefont {Evertz}, \citenamefont {Feiguin},
  \citenamefont {Freire}, \citenamefont {Fuchs}, \citenamefont {Gamper},
  \citenamefont {Gukelberger}, \citenamefont {Gull}, \citenamefont {Guertler},
  \citenamefont {Hehn}, \citenamefont {Igarashi}, \citenamefont {Isakov},
  \citenamefont {Koop}, \citenamefont {Ma}, \citenamefont {Mates},
  \citenamefont {Matsuo}, \citenamefont {Parcollet}, \citenamefont
  {Pawłowski}, \citenamefont {Picon}, \citenamefont {Pollet}, \citenamefont
  {Santos}, \citenamefont {Scarola}, \citenamefont {Schollwöck}, \citenamefont
  {Silva}, \citenamefont {Surer}, \citenamefont {Todo}, \citenamefont {Trebst},
  \citenamefont {Troyer}, \citenamefont {Wall}, \citenamefont {Werner},\ and\
  \citenamefont {Wessel}}]{ALPS2}%
  \BibitemOpen
  \bibfield  {author} {\bibinfo {author} {\bibfnamefont {B.}~\bibnamefont
  {Bauer}}, \bibinfo {author} {\bibfnamefont {L.~D.}\ \bibnamefont {Carr}},
  \bibinfo {author} {\bibfnamefont {H.~G.}\ \bibnamefont {Evertz}}, \bibinfo
  {author} {\bibfnamefont {A.}~\bibnamefont {Feiguin}}, \bibinfo {author}
  {\bibfnamefont {J.}~\bibnamefont {Freire}}, \bibinfo {author} {\bibfnamefont
  {S.}~\bibnamefont {Fuchs}}, \bibinfo {author} {\bibfnamefont
  {L.}~\bibnamefont {Gamper}}, \bibinfo {author} {\bibfnamefont
  {J.}~\bibnamefont {Gukelberger}}, \bibinfo {author} {\bibfnamefont
  {E.}~\bibnamefont {Gull}}, \bibinfo {author} {\bibfnamefont {S.}~\bibnamefont
  {Guertler}}, \bibinfo {author} {\bibfnamefont {A.}~\bibnamefont {Hehn}},
  \bibinfo {author} {\bibfnamefont {R.}~\bibnamefont {Igarashi}}, \bibinfo
  {author} {\bibfnamefont {S.~V.}\ \bibnamefont {Isakov}}, \bibinfo {author}
  {\bibfnamefont {D.}~\bibnamefont {Koop}}, \bibinfo {author} {\bibfnamefont
  {P.~N.}\ \bibnamefont {Ma}}, \bibinfo {author} {\bibfnamefont
  {P.}~\bibnamefont {Mates}}, \bibinfo {author} {\bibfnamefont
  {H.}~\bibnamefont {Matsuo}}, \bibinfo {author} {\bibfnamefont
  {O.}~\bibnamefont {Parcollet}}, \bibinfo {author} {\bibfnamefont
  {G.}~\bibnamefont {Pawłowski}}, \bibinfo {author} {\bibfnamefont {J.~D.}\
  \bibnamefont {Picon}}, \bibinfo {author} {\bibfnamefont {L.}~\bibnamefont
  {Pollet}}, \bibinfo {author} {\bibfnamefont {E.}~\bibnamefont {Santos}},
  \bibinfo {author} {\bibfnamefont {V.~W.}\ \bibnamefont {Scarola}}, \bibinfo
  {author} {\bibfnamefont {U.}~\bibnamefont {Schollwöck}}, \bibinfo {author}
  {\bibfnamefont {C.}~\bibnamefont {Silva}}, \bibinfo {author} {\bibfnamefont
  {B.}~\bibnamefont {Surer}}, \bibinfo {author} {\bibfnamefont
  {S.}~\bibnamefont {Todo}}, \bibinfo {author} {\bibfnamefont {S.}~\bibnamefont
  {Trebst}}, \bibinfo {author} {\bibfnamefont {M.}~\bibnamefont {Troyer}},
  \bibinfo {author} {\bibfnamefont {M.~L.}\ \bibnamefont {Wall}}, \bibinfo
  {author} {\bibfnamefont {P.}~\bibnamefont {Werner}}, \ and\ \bibinfo {author}
  {\bibfnamefont {S.}~\bibnamefont {Wessel}},\ }\href {\doibase
  10.1088/1742-5468/2011/05/P05001} {\bibfield  {journal} {\bibinfo  {journal}
  {Journal of Statistical Mechanics: Theory and Experiment}\ }\textbf {\bibinfo
  {volume} {2011}},\ \bibinfo {pages} {P05001} (\bibinfo {year}
  {2011})}\BibitemShut {NoStop}%
\bibitem [{\citenamefont {Huang}\ \emph {et~al.}(2015)\citenamefont {Huang},
  \citenamefont {Wang}, \citenamefont {Meng}, \citenamefont {Du}, \citenamefont
  {Werner},\ and\ \citenamefont {Dai}}]{Huang15}%
  \BibitemOpen
  \bibfield  {author} {\bibinfo {author} {\bibfnamefont {L.}~\bibnamefont
  {Huang}}, \bibinfo {author} {\bibfnamefont {Y.}~\bibnamefont {Wang}},
  \bibinfo {author} {\bibfnamefont {Z.~Y.}\ \bibnamefont {Meng}}, \bibinfo
  {author} {\bibfnamefont {L.}~\bibnamefont {Du}}, \bibinfo {author}
  {\bibfnamefont {P.}~\bibnamefont {Werner}}, \ and\ \bibinfo {author}
  {\bibfnamefont {X.}~\bibnamefont {Dai}},\ }\href {\doibase
  10.1016/j.cpc.2015.04.020} {\bibfield  {journal} {\bibinfo  {journal} {Comp.
  Phys. Comm.}\ }\textbf {\bibinfo {volume} {195}},\ \bibinfo {pages} {140}
  (\bibinfo {year} {2015})}\BibitemShut {NoStop}%
\bibitem [{\citenamefont {Huang}(2017)}]{Huang17}%
  \BibitemOpen
  \bibfield  {author} {\bibinfo {author} {\bibfnamefont {L.}~\bibnamefont
  {Huang}},\ }\href {\doibase 10.1016/j.cpc.2017.08.026} {\bibfield  {journal}
  {\bibinfo  {journal} {Comp. Phys. Comm.}\ }\textbf {\bibinfo {volume}
  {221}},\ \bibinfo {pages} {423} (\bibinfo {year} {2017})}\BibitemShut
  {NoStop}%
\bibitem [{\citenamefont {Hafermann}\ \emph {et~al.}(2013)\citenamefont
  {Hafermann}, \citenamefont {Werner},\ and\ \citenamefont
  {Gull}}]{Hafermann13}%
  \BibitemOpen
  \bibfield  {author} {\bibinfo {author} {\bibfnamefont {H.}~\bibnamefont
  {Hafermann}}, \bibinfo {author} {\bibfnamefont {P.}~\bibnamefont {Werner}}, \
  and\ \bibinfo {author} {\bibfnamefont {E.}~\bibnamefont {Gull}},\ }\href
  {\doibase http://dx.doi.org/10.1016/j.cpc.2012.12.013} {\bibfield  {journal}
  {\bibinfo  {journal} {Computer Physics Communications}\ }\textbf {\bibinfo
  {volume} {184}},\ \bibinfo {pages} {1280 } (\bibinfo {year}
  {2013})}\BibitemShut {NoStop}%
\bibitem [{\citenamefont {Hafermann}\ \emph {et~al.}(2012)\citenamefont
  {Hafermann}, \citenamefont {Patton},\ and\ \citenamefont
  {Werner}}]{Hafermann12}%
  \BibitemOpen
  \bibfield  {author} {\bibinfo {author} {\bibfnamefont {H.}~\bibnamefont
  {Hafermann}}, \bibinfo {author} {\bibfnamefont {K.~R.}\ \bibnamefont
  {Patton}}, \ and\ \bibinfo {author} {\bibfnamefont {P.}~\bibnamefont
  {Werner}},\ }\href {\doibase 10.1103/PhysRevB.85.205106} {\bibfield
  {journal} {\bibinfo  {journal} {Phys. Rev. B}\ }\textbf {\bibinfo {volume}
  {85}},\ \bibinfo {pages} {205106} (\bibinfo {year} {2012})}\BibitemShut
  {NoStop}%
\bibitem [{Note2()}]{Note2}%
  \BibitemOpen
  \bibinfo {note} {Here we ignore the possibility of mixing of previous and
  current iterative solutions, since we are interested in the fundamental
  aspects of the stability of DMFT solutions.}\BibitemShut {Stop}%
\bibitem [{\citenamefont {Rubtsov}\ \emph {et~al.}(2008)\citenamefont
  {Rubtsov}, \citenamefont {Katsnelson},\ and\ \citenamefont
  {Lichtenstein}}]{Rubtsov08}%
  \BibitemOpen
  \bibfield  {author} {\bibinfo {author} {\bibfnamefont {A.~N.}\ \bibnamefont
  {Rubtsov}}, \bibinfo {author} {\bibfnamefont {M.~I.}\ \bibnamefont
  {Katsnelson}}, \ and\ \bibinfo {author} {\bibfnamefont {A.~I.}\ \bibnamefont
  {Lichtenstein}},\ }\href {\doibase 10.1103/PhysRevB.77.033101} {\bibfield
  {journal} {\bibinfo  {journal} {Phys. Rev. B}\ }\textbf {\bibinfo {volume}
  {77}},\ \bibinfo {pages} {033101} (\bibinfo {year} {2008})}\BibitemShut
  {NoStop}%
\bibitem [{\citenamefont {Hafermann}(2010)}]{Hafermannphd}%
  \BibitemOpen
  \bibfield  {author} {\bibinfo {author} {\bibfnamefont {H.}~\bibnamefont
  {Hafermann}},\ }\emph {\bibinfo {title} {Numerical Approaches to Spatial
  Correlations in Strongly Interacting Fermion Systems}},\ \href@noop {} {Ph.D.
  thesis},\ \bibinfo  {school} {University of Hamburg} (\bibinfo {year}
  {2010})\BibitemShut {NoStop}%
\bibitem [{\citenamefont {Rohringer}\ \emph {et~al.}(2018)\citenamefont
  {Rohringer}, \citenamefont {Hafermann}, \citenamefont {Toschi}, \citenamefont
  {Katanin}, \citenamefont {Antipov}, \citenamefont {Katsnelson}, \citenamefont
  {Lichtenstein}, \citenamefont {Rubtsov},\ and\ \citenamefont
  {Held}}]{Rohringer18}%
  \BibitemOpen
  \bibfield  {author} {\bibinfo {author} {\bibfnamefont {G.}~\bibnamefont
  {Rohringer}}, \bibinfo {author} {\bibfnamefont {H.}~\bibnamefont
  {Hafermann}}, \bibinfo {author} {\bibfnamefont {A.}~\bibnamefont {Toschi}},
  \bibinfo {author} {\bibfnamefont {A.~A.}\ \bibnamefont {Katanin}}, \bibinfo
  {author} {\bibfnamefont {A.~E.}\ \bibnamefont {Antipov}}, \bibinfo {author}
  {\bibfnamefont {M.~I.}\ \bibnamefont {Katsnelson}}, \bibinfo {author}
  {\bibfnamefont {A.~I.}\ \bibnamefont {Lichtenstein}}, \bibinfo {author}
  {\bibfnamefont {A.~N.}\ \bibnamefont {Rubtsov}}, \ and\ \bibinfo {author}
  {\bibfnamefont {K.}~\bibnamefont {Held}},\ }\href {\doibase
  10.1103/RevModPhys.90.025003} {\bibfield  {journal} {\bibinfo  {journal}
  {Rev. Mod. Phys.}\ }\textbf {\bibinfo {volume} {90}},\ \bibinfo {pages}
  {025003} (\bibinfo {year} {2018})}\BibitemShut {NoStop}%
\bibitem [{\citenamefont {van Loon}\ \emph {et~al.}(2015)\citenamefont {van
  Loon}, \citenamefont {Hafermann}, \citenamefont {Lichtenstein},\ and\
  \citenamefont {Katsnelson}}]{vanLoon15}%
  \BibitemOpen
  \bibfield  {author} {\bibinfo {author} {\bibfnamefont {E.~G. C.~P.}\
  \bibnamefont {van Loon}}, \bibinfo {author} {\bibfnamefont {H.}~\bibnamefont
  {Hafermann}}, \bibinfo {author} {\bibfnamefont {A.~I.}\ \bibnamefont
  {Lichtenstein}}, \ and\ \bibinfo {author} {\bibfnamefont {M.~I.}\
  \bibnamefont {Katsnelson}},\ }\href {\doibase 10.1103/PhysRevB.92.085106}
  {\bibfield  {journal} {\bibinfo  {journal} {Phys. Rev. B}\ }\textbf {\bibinfo
  {volume} {92}},\ \bibinfo {pages} {085106} (\bibinfo {year}
  {2015})}\BibitemShut {NoStop}%
\bibitem [{\citenamefont {Kotliar}(1999)}]{Kotliar99}%
  \BibitemOpen
  \bibfield  {author} {\bibinfo {author} {\bibfnamefont {G.}~\bibnamefont
  {Kotliar}},\ }\href {\doibase 10.1007/s100510050914} {\bibfield  {journal}
  {\bibinfo  {journal} {The European Physical Journal B - Condensed Matter and
  Complex Systems}\ }\textbf {\bibinfo {volume} {11}},\ \bibinfo {pages} {27}
  (\bibinfo {year} {1999})}\BibitemShut {NoStop}%
\bibitem [{\citenamefont {Kotliar}\ \emph {et~al.}(2000)\citenamefont
  {Kotliar}, \citenamefont {Lange},\ and\ \citenamefont
  {Rozenberg}}]{Kotliar00}%
  \BibitemOpen
  \bibfield  {author} {\bibinfo {author} {\bibfnamefont {G.}~\bibnamefont
  {Kotliar}}, \bibinfo {author} {\bibfnamefont {E.}~\bibnamefont {Lange}}, \
  and\ \bibinfo {author} {\bibfnamefont {M.~J.}\ \bibnamefont {Rozenberg}},\
  }\href {\doibase 10.1103/PhysRevLett.84.5180} {\bibfield  {journal} {\bibinfo
   {journal} {Phys. Rev. Lett.}\ }\textbf {\bibinfo {volume} {84}},\ \bibinfo
  {pages} {5180} (\bibinfo {year} {2000})}\BibitemShut {NoStop}%
\bibitem [{\citenamefont {Astretsov}\ \emph {et~al.}(2020)\citenamefont
  {Astretsov}, \citenamefont {Rohringer},\ and\ \citenamefont
  {Rubtsov}}]{Astretsov20}%
  \BibitemOpen
  \bibfield  {author} {\bibinfo {author} {\bibfnamefont {G.~V.}\ \bibnamefont
  {Astretsov}}, \bibinfo {author} {\bibfnamefont {G.}~\bibnamefont
  {Rohringer}}, \ and\ \bibinfo {author} {\bibfnamefont {A.~N.}\ \bibnamefont
  {Rubtsov}},\ }\href {\doibase 10.1103/PhysRevB.101.075109} {\bibfield
  {journal} {\bibinfo  {journal} {Phys. Rev. B}\ }\textbf {\bibinfo {volume}
  {101}},\ \bibinfo {pages} {075109} (\bibinfo {year} {2020})}\BibitemShut
  {NoStop}%
\bibitem [{\citenamefont {Kotliar}\ \emph {et~al.}(2002)\citenamefont
  {Kotliar}, \citenamefont {Murthy},\ and\ \citenamefont
  {Rozenberg}}]{Kotliar02}%
  \BibitemOpen
  \bibfield  {author} {\bibinfo {author} {\bibfnamefont {G.}~\bibnamefont
  {Kotliar}}, \bibinfo {author} {\bibfnamefont {S.}~\bibnamefont {Murthy}}, \
  and\ \bibinfo {author} {\bibfnamefont {M.~J.}\ \bibnamefont {Rozenberg}},\
  }\href {\doibase 10.1103/PhysRevLett.89.046401} {\bibfield  {journal}
  {\bibinfo  {journal} {Phys. Rev. Lett.}\ }\textbf {\bibinfo {volume} {89}},\
  \bibinfo {pages} {046401} (\bibinfo {year} {2002})}\BibitemShut {NoStop}%
\bibitem [{\citenamefont {Hafermann}\ \emph {et~al.}(2014)\citenamefont
  {Hafermann}, \citenamefont {van Loon}, \citenamefont {Katsnelson},
  \citenamefont {Lichtenstein},\ and\ \citenamefont
  {Parcollet}}]{Hafermann14b}%
  \BibitemOpen
  \bibfield  {author} {\bibinfo {author} {\bibfnamefont {H.}~\bibnamefont
  {Hafermann}}, \bibinfo {author} {\bibfnamefont {E.~G. C.~P.}\ \bibnamefont
  {van Loon}}, \bibinfo {author} {\bibfnamefont {M.~I.}\ \bibnamefont
  {Katsnelson}}, \bibinfo {author} {\bibfnamefont {A.~I.}\ \bibnamefont
  {Lichtenstein}}, \ and\ \bibinfo {author} {\bibfnamefont {O.}~\bibnamefont
  {Parcollet}},\ }\href {\doibase 10.1103/PhysRevB.90.235105} {\bibfield
  {journal} {\bibinfo  {journal} {Phys. Rev. B}\ }\textbf {\bibinfo {volume}
  {90}},\ \bibinfo {pages} {235105} (\bibinfo {year} {2014})}\BibitemShut
  {NoStop}%
\bibitem [{\citenamefont {Werner}\ \emph {et~al.}(2008)\citenamefont {Werner},
  \citenamefont {Gull}, \citenamefont {Troyer},\ and\ \citenamefont
  {Millis}}]{Werner08}%
  \BibitemOpen
  \bibfield  {author} {\bibinfo {author} {\bibfnamefont {P.}~\bibnamefont
  {Werner}}, \bibinfo {author} {\bibfnamefont {E.}~\bibnamefont {Gull}},
  \bibinfo {author} {\bibfnamefont {M.}~\bibnamefont {Troyer}}, \ and\ \bibinfo
  {author} {\bibfnamefont {A.~J.}\ \bibnamefont {Millis}},\ }\href {\doibase
  10.1103/PhysRevLett.101.166405} {\bibfield  {journal} {\bibinfo  {journal}
  {Phys. Rev. Lett.}\ }\textbf {\bibinfo {volume} {101}},\ \bibinfo {pages}
  {166405} (\bibinfo {year} {2008})}\BibitemShut {NoStop}%
\bibitem [{\citenamefont {Haule}\ and\ \citenamefont
  {Kotliar}(2009)}]{Haule09}%
  \BibitemOpen
  \bibfield  {author} {\bibinfo {author} {\bibfnamefont {K.}~\bibnamefont
  {Haule}}\ and\ \bibinfo {author} {\bibfnamefont {G.}~\bibnamefont
  {Kotliar}},\ }\href {\doibase 10.1088/1367-2630/11/2/025021} {\bibfield
  {journal} {\bibinfo  {journal} {New Journal of Physics}\ }\textbf {\bibinfo
  {volume} {11}},\ \bibinfo {pages} {025021} (\bibinfo {year}
  {2009})}\BibitemShut {NoStop}%
\bibitem [{\citenamefont {de' Medici}\ \emph {et~al.}(2009)\citenamefont {de'
  Medici}, \citenamefont {Hassan}, \citenamefont {Capone},\ and\ \citenamefont
  {Dai}}]{deMedici09}%
  \BibitemOpen
  \bibfield  {author} {\bibinfo {author} {\bibfnamefont {L.}~\bibnamefont {de'
  Medici}}, \bibinfo {author} {\bibfnamefont {S.~R.}\ \bibnamefont {Hassan}},
  \bibinfo {author} {\bibfnamefont {M.}~\bibnamefont {Capone}}, \ and\ \bibinfo
  {author} {\bibfnamefont {X.}~\bibnamefont {Dai}},\ }\href {\doibase
  10.1103/PhysRevLett.102.126401} {\bibfield  {journal} {\bibinfo  {journal}
  {Phys. Rev. Lett.}\ }\textbf {\bibinfo {volume} {102}},\ \bibinfo {pages}
  {126401} (\bibinfo {year} {2009})}\BibitemShut {NoStop}%
\bibitem [{\citenamefont {de' Medici}\ \emph {et~al.}(2011)\citenamefont {de'
  Medici}, \citenamefont {Mravlje},\ and\ \citenamefont
  {Georges}}]{deMedici11}%
  \BibitemOpen
  \bibfield  {author} {\bibinfo {author} {\bibfnamefont {L.}~\bibnamefont {de'
  Medici}}, \bibinfo {author} {\bibfnamefont {J.}~\bibnamefont {Mravlje}}, \
  and\ \bibinfo {author} {\bibfnamefont {A.}~\bibnamefont {Georges}},\ }\href
  {\doibase 10.1103/PhysRevLett.107.256401} {\bibfield  {journal} {\bibinfo
  {journal} {Phys. Rev. Lett.}\ }\textbf {\bibinfo {volume} {107}},\ \bibinfo
  {pages} {256401} (\bibinfo {year} {2011})}\BibitemShut {NoStop}%
\bibitem [{\citenamefont {Werner}\ \emph {et~al.}(2012)\citenamefont {Werner},
  \citenamefont {Casula}, \citenamefont {Miyake}, \citenamefont {Aryasetiawan},
  \citenamefont {Millis},\ and\ \citenamefont {Biermann}}]{Werner12}%
  \BibitemOpen
  \bibfield  {author} {\bibinfo {author} {\bibfnamefont {P.}~\bibnamefont
  {Werner}}, \bibinfo {author} {\bibfnamefont {M.}~\bibnamefont {Casula}},
  \bibinfo {author} {\bibfnamefont {T.}~\bibnamefont {Miyake}}, \bibinfo
  {author} {\bibfnamefont {F.}~\bibnamefont {Aryasetiawan}}, \bibinfo {author}
  {\bibfnamefont {A.~J.}\ \bibnamefont {Millis}}, \ and\ \bibinfo {author}
  {\bibfnamefont {S.}~\bibnamefont {Biermann}},\ }\href@noop {} {\bibfield
  {journal} {\bibinfo  {journal} {Nature Physics}\ }\textbf {\bibinfo {volume}
  {8}},\ \bibinfo {pages} {331} (\bibinfo {year} {2012})}\BibitemShut {NoStop}%
\bibitem [{\citenamefont {Stadler}\ \emph {et~al.}(2015)\citenamefont
  {Stadler}, \citenamefont {Yin}, \citenamefont {von Delft}, \citenamefont
  {Kotliar},\ and\ \citenamefont {Weichselbaum}}]{Stadler15}%
  \BibitemOpen
  \bibfield  {author} {\bibinfo {author} {\bibfnamefont {K.~M.}\ \bibnamefont
  {Stadler}}, \bibinfo {author} {\bibfnamefont {Z.~P.}\ \bibnamefont {Yin}},
  \bibinfo {author} {\bibfnamefont {J.}~\bibnamefont {von Delft}}, \bibinfo
  {author} {\bibfnamefont {G.}~\bibnamefont {Kotliar}}, \ and\ \bibinfo
  {author} {\bibfnamefont {A.}~\bibnamefont {Weichselbaum}},\ }\href {\doibase
  10.1103/PhysRevLett.115.136401} {\bibfield  {journal} {\bibinfo  {journal}
  {Phys. Rev. Lett.}\ }\textbf {\bibinfo {volume} {115}},\ \bibinfo {pages}
  {136401} (\bibinfo {year} {2015})}\BibitemShut {NoStop}%
\bibitem [{\citenamefont {de' Medici}(2017)}]{Medici17}%
  \BibitemOpen
  \bibfield  {author} {\bibinfo {author} {\bibfnamefont {L.}~\bibnamefont {de'
  Medici}},\ }\href {\doibase 10.1103/PhysRevLett.118.167003} {\bibfield
  {journal} {\bibinfo  {journal} {Phys. Rev. Lett.}\ }\textbf {\bibinfo
  {volume} {118}},\ \bibinfo {pages} {167003} (\bibinfo {year}
  {2017})}\BibitemShut {NoStop}%
\bibitem [{\citenamefont {Villar~Arribi}\ and\ \citenamefont {de'
  Medici}(2018)}]{Villar18}%
  \BibitemOpen
  \bibfield  {author} {\bibinfo {author} {\bibfnamefont {P.}~\bibnamefont
  {Villar~Arribi}}\ and\ \bibinfo {author} {\bibfnamefont {L.}~\bibnamefont
  {de' Medici}},\ }\href {\doibase 10.1103/PhysRevLett.121.197001} {\bibfield
  {journal} {\bibinfo  {journal} {Phys. Rev. Lett.}\ }\textbf {\bibinfo
  {volume} {121}},\ \bibinfo {pages} {197001} (\bibinfo {year}
  {2018})}\BibitemShut {NoStop}%
\bibitem [{\citenamefont {Grilli}\ \emph {et~al.}(1991)\citenamefont {Grilli},
  \citenamefont {Raimondi}, \citenamefont {Castellani}, \citenamefont
  {Di~Castro},\ and\ \citenamefont {Kotliar}}]{Grilli91}%
  \BibitemOpen
  \bibfield  {author} {\bibinfo {author} {\bibfnamefont {M.}~\bibnamefont
  {Grilli}}, \bibinfo {author} {\bibfnamefont {R.}~\bibnamefont {Raimondi}},
  \bibinfo {author} {\bibfnamefont {C.}~\bibnamefont {Castellani}}, \bibinfo
  {author} {\bibfnamefont {C.}~\bibnamefont {Di~Castro}}, \ and\ \bibinfo
  {author} {\bibfnamefont {G.}~\bibnamefont {Kotliar}},\ }\href {\doibase
  10.1103/PhysRevLett.67.259} {\bibfield  {journal} {\bibinfo  {journal} {Phys.
  Rev. Lett.}\ }\textbf {\bibinfo {volume} {67}},\ \bibinfo {pages} {259}
  (\bibinfo {year} {1991})}\BibitemShut {NoStop}%
\bibitem [{\citenamefont {Majumdar}\ and\ \citenamefont
  {Krishnamurthy}(1994)}]{Majumdar94}%
  \BibitemOpen
  \bibfield  {author} {\bibinfo {author} {\bibfnamefont {P.}~\bibnamefont
  {Majumdar}}\ and\ \bibinfo {author} {\bibfnamefont {H.~R.}\ \bibnamefont
  {Krishnamurthy}},\ }\href {\doibase 10.1103/PhysRevLett.73.1525} {\bibfield
  {journal} {\bibinfo  {journal} {Phys. Rev. Lett.}\ }\textbf {\bibinfo
  {volume} {73}},\ \bibinfo {pages} {1525} (\bibinfo {year}
  {1994})}\BibitemShut {NoStop}%
\bibitem [{\citenamefont {Tandon}\ \emph {et~al.}(1999)\citenamefont {Tandon},
  \citenamefont {Wang},\ and\ \citenamefont {Kotliar}}]{Tandon99}%
  \BibitemOpen
  \bibfield  {author} {\bibinfo {author} {\bibfnamefont {A.}~\bibnamefont
  {Tandon}}, \bibinfo {author} {\bibfnamefont {Z.}~\bibnamefont {Wang}}, \ and\
  \bibinfo {author} {\bibfnamefont {G.}~\bibnamefont {Kotliar}},\ }\href
  {\doibase 10.1103/PhysRevLett.83.2046} {\bibfield  {journal} {\bibinfo
  {journal} {Phys. Rev. Lett.}\ }\textbf {\bibinfo {volume} {83}},\ \bibinfo
  {pages} {2046} (\bibinfo {year} {1999})}\BibitemShut {NoStop}%
\bibitem [{\citenamefont {Held}\ \emph {et~al.}(2001)\citenamefont {Held},
  \citenamefont {McMahan},\ and\ \citenamefont {Scalettar}}]{Held01}%
  \BibitemOpen
  \bibfield  {author} {\bibinfo {author} {\bibfnamefont {K.}~\bibnamefont
  {Held}}, \bibinfo {author} {\bibfnamefont {A.~K.}\ \bibnamefont {McMahan}}, \
  and\ \bibinfo {author} {\bibfnamefont {R.~T.}\ \bibnamefont {Scalettar}},\
  }\href {\doibase 10.1103/PhysRevLett.87.276404} {\bibfield  {journal}
  {\bibinfo  {journal} {Phys. Rev. Lett.}\ }\textbf {\bibinfo {volume} {87}},\
  \bibinfo {pages} {276404} (\bibinfo {year} {2001})}\BibitemShut {NoStop}%
\bibitem [{\citenamefont {Capone}\ \emph {et~al.}(2004)\citenamefont {Capone},
  \citenamefont {Sangiovanni}, \citenamefont {Castellani}, \citenamefont
  {Di~Castro},\ and\ \citenamefont {Grilli}}]{Capone04}%
  \BibitemOpen
  \bibfield  {author} {\bibinfo {author} {\bibfnamefont {M.}~\bibnamefont
  {Capone}}, \bibinfo {author} {\bibfnamefont {G.}~\bibnamefont {Sangiovanni}},
  \bibinfo {author} {\bibfnamefont {C.}~\bibnamefont {Castellani}}, \bibinfo
  {author} {\bibfnamefont {C.}~\bibnamefont {Di~Castro}}, \ and\ \bibinfo
  {author} {\bibfnamefont {M.}~\bibnamefont {Grilli}},\ }\href {\doibase
  10.1103/PhysRevLett.92.106401} {\bibfield  {journal} {\bibinfo  {journal}
  {Phys. Rev. Lett.}\ }\textbf {\bibinfo {volume} {92}},\ \bibinfo {pages}
  {106401} (\bibinfo {year} {2004})}\BibitemShut {NoStop}%
\bibitem [{\citenamefont {Aichhorn}\ \emph {et~al.}(2007)\citenamefont
  {Aichhorn}, \citenamefont {Arrigoni}, \citenamefont {Potthoff},\ and\
  \citenamefont {Hanke}}]{Aichhorn07}%
  \BibitemOpen
  \bibfield  {author} {\bibinfo {author} {\bibfnamefont {M.}~\bibnamefont
  {Aichhorn}}, \bibinfo {author} {\bibfnamefont {E.}~\bibnamefont {Arrigoni}},
  \bibinfo {author} {\bibfnamefont {M.}~\bibnamefont {Potthoff}}, \ and\
  \bibinfo {author} {\bibfnamefont {W.}~\bibnamefont {Hanke}},\ }\href
  {\doibase 10.1103/PhysRevB.76.224509} {\bibfield  {journal} {\bibinfo
  {journal} {Phys. Rev. B}\ }\textbf {\bibinfo {volume} {76}},\ \bibinfo
  {pages} {224509} (\bibinfo {year} {2007})}\BibitemShut {NoStop}%
\bibitem [{\citenamefont {Lupi}\ \emph {et~al.}(2010)\citenamefont {Lupi},
  \citenamefont {Baldassarre}, \citenamefont {Mansart}, \citenamefont
  {Perucchi}, \citenamefont {Barinov}, \citenamefont {Dudin}, \citenamefont
  {Papalazarou}, \citenamefont {Rodolakis}, \citenamefont {Rueff},
  \citenamefont {Iti{\'e}} \emph {et~al.}}]{Lupi10}%
  \BibitemOpen
  \bibfield  {author} {\bibinfo {author} {\bibfnamefont {S.}~\bibnamefont
  {Lupi}}, \bibinfo {author} {\bibfnamefont {L.}~\bibnamefont {Baldassarre}},
  \bibinfo {author} {\bibfnamefont {B.}~\bibnamefont {Mansart}}, \bibinfo
  {author} {\bibfnamefont {A.}~\bibnamefont {Perucchi}}, \bibinfo {author}
  {\bibfnamefont {A.}~\bibnamefont {Barinov}}, \bibinfo {author} {\bibfnamefont
  {P.}~\bibnamefont {Dudin}}, \bibinfo {author} {\bibfnamefont
  {E.}~\bibnamefont {Papalazarou}}, \bibinfo {author} {\bibfnamefont
  {F.}~\bibnamefont {Rodolakis}}, \bibinfo {author} {\bibfnamefont {J.-P.}\
  \bibnamefont {Rueff}}, \bibinfo {author} {\bibfnamefont {J.-P.}\ \bibnamefont
  {Iti{\'e}}},  \emph {et~al.},\ }\href@noop {} {\bibfield  {journal} {\bibinfo
   {journal} {Nature communications}\ }\textbf {\bibinfo {volume} {1}},\
  \bibinfo {pages} {105} (\bibinfo {year} {2010})}\BibitemShut {NoStop}%
\bibitem [{\citenamefont {Otsuki}\ \emph {et~al.}(2014)\citenamefont {Otsuki},
  \citenamefont {Hafermann},\ and\ \citenamefont {Lichtenstein}}]{Otsuki14}%
  \BibitemOpen
  \bibfield  {author} {\bibinfo {author} {\bibfnamefont {J.}~\bibnamefont
  {Otsuki}}, \bibinfo {author} {\bibfnamefont {H.}~\bibnamefont {Hafermann}}, \
  and\ \bibinfo {author} {\bibfnamefont {A.~I.}\ \bibnamefont {Lichtenstein}},\
  }\href {\doibase 10.1103/PhysRevB.90.235132} {\bibfield  {journal} {\bibinfo
  {journal} {Phys. Rev. B}\ }\textbf {\bibinfo {volume} {90}},\ \bibinfo
  {pages} {235132} (\bibinfo {year} {2014})}\BibitemShut {NoStop}%
\bibitem [{\citenamefont {Yee}\ and\ \citenamefont {Balents}(2015)}]{Yee15}%
  \BibitemOpen
  \bibfield  {author} {\bibinfo {author} {\bibfnamefont {C.-H.}\ \bibnamefont
  {Yee}}\ and\ \bibinfo {author} {\bibfnamefont {L.}~\bibnamefont {Balents}},\
  }\href {\doibase 10.1103/PhysRevX.5.021007} {\bibfield  {journal} {\bibinfo
  {journal} {Phys. Rev. X}\ }\textbf {\bibinfo {volume} {5}},\ \bibinfo {pages}
  {021007} (\bibinfo {year} {2015})}\BibitemShut {NoStop}%
\bibitem [{\citenamefont {Logan}\ and\ \citenamefont {Galpin}(2015)}]{Logan15}%
  \BibitemOpen
  \bibfield  {author} {\bibinfo {author} {\bibfnamefont {D.~E.}\ \bibnamefont
  {Logan}}\ and\ \bibinfo {author} {\bibfnamefont {M.~R.}\ \bibnamefont
  {Galpin}},\ }\href {\doibase 10.1088/0953-8984/28/2/025601} {\bibfield
  {journal} {\bibinfo  {journal} {Journal of Physics: Condensed Matter}\
  }\textbf {\bibinfo {volume} {28}},\ \bibinfo {pages} {025601} (\bibinfo
  {year} {2015})}\BibitemShut {NoStop}%
\bibitem [{\citenamefont {Sen}\ \emph {et~al.}(2020)\citenamefont {Sen},
  \citenamefont {Wong},\ and\ \citenamefont {Mitchell}}]{Sen20}%
  \BibitemOpen
  \bibfield  {author} {\bibinfo {author} {\bibfnamefont {S.}~\bibnamefont
  {Sen}}, \bibinfo {author} {\bibfnamefont {P.~J.}\ \bibnamefont {Wong}}, \
  and\ \bibinfo {author} {\bibfnamefont {A.~K.}\ \bibnamefont {Mitchell}},\
  }\href@noop {} {\enquote {\bibinfo {title} {The {Mott} transition as a
  topological phase transition},}\ } (\bibinfo {year} {2020}),\ \Eprint
  {http://arxiv.org/abs/2001.10526} {arXiv:2001.10526 [cond-mat.str-el]}
  \BibitemShut {NoStop}%
\end{thebibliography}%

\clearpage 

\section{Supplemental material: Conventions and detailed derivations}

\subsection{ The auxiliary impurity model: thermodynamic potential and its derivatives}

The auxiliary impurity model is a central object in DMFT.
In the action formalism, it is defined as 
\begin{align}
    S_{\rm imp} &=  T \sum_\nu c_{\nu,\sigma}^* (\Delta_\nu-i \nu_n) c_{\nu,\sigma} + U \int_0^\beta d\tau \, \, n_{\up}(\tau)n_{\dn}(\tau),\notag
\end{align}
where $c_\sigma(\tau),c^*_\sigma(\tau)$ are Grassmann numbers, $c_{\nu,\sigma},c^*_{\nu,\sigma}$ are their Fourier transform, and $n_{\sigma}(\tau)=c^*_\sigma(\tau)c_\sigma(\tau)$. The corresponding thermodynamic potential reads  
\begin{align}
\Omega_{\rm imp}&=-T\ln Z,\notag \\
    Z&= \int d[c^*,c] \exp(-S_{\rm imp}).
\end{align}
Accordingly, the first derivative of the thermodynamic potential with respect to the hybridization yields the single-particle Green's function of the impurity,
\begin{align}
    \frac{\delta \Omega_{\rm imp}}{\delta \Delta_\nu} 
    = -T\av{cc^*}_\nu \equiv T g_\nu, 
\end{align}
and the second derivative yields
\begin{align}
    \frac{\delta g_\nu}{\delta \Delta_{\nu'}} 
    &=\beta \frac{\delta^2 \Omega_{\rm imp}}{\delta \Delta_\nu \delta \Delta_{\nu'}} \notag\\
    &=g_\nu g_{\nu'}- \frac{1}{Z} \int d[c^*,c] (c^* c)_{\nu}(c^* c)_{\nu'} \exp(-S)\notag\\
    &= -T \av{cc^* cc^*}_{\nu\nu',\omega=0} + g_\nu g_{\nu'}.
\end{align}
We relate the two-particle Green's function $\av{cc^* cc^*}$ to the local vertex $F^{\rm loc}_{\nu\nu',\omega}$ by
\begin{align}
    T \av{cc^* cc^*}_{\nu\nu',\omega=0}\!=\!g_\nu g_{\nu'}\!-\!g_\nu^2 \delta_{\nu \nu'}\!-\!Tg_\nu^2 F^{\rm loc}_{\nu\nu',\omega=0} g_{\nu'}^2,
\end{align}
which yields 
\begin{align}
    \frac{\delta g_\nu}{\delta \Delta_{\nu'}}=g_\nu^2 \left(\delta_{\nu \nu'}+ TF^{\rm loc}_{\nu\nu',\omega=0} g_{\nu'}^2\right)=-T\chi^{\rm loc}_{\nu\nu'\omega=0}.\label{dGloc}
\end{align}
where $\chi^{\rm loc}_{\nu\nu'\omega}$ is the local charge susceptibility. Comparing this to
\begin{align}
\frac{\delta g_\nu}{\delta \Delta_{\nu'}}=g_\nu^2\left( \delta_{\nu \nu'}+\frac{\delta\Sigma_\nu}{\delta\Delta_{\nu'}}\right)
\end{align}
we finally obtain the useful relation
\begin{align}
\frac{\delta\Sigma _{\nu }}{\delta\Delta _{\nu ^{\prime }}}=T F^{\rm loc}_{\nu \nu'}g_{\nu ^{\prime }}^{2}.\label{dSigma}
\end{align}

\subsection{The Jacobian of DMFT self-consistency}

We analyze the stability of DMFT iterations in terms of $\Delta$, the usual input of impurity solvers.
In the DMFT self-consistency condition $\sum_\kv G_{\kv,\nu} =g_\nu$, $\Delta$ appears explicitly on the right-hand side, since $g^{-1}=i\nu-\Delta_\nu-\Sigma$, but it also appears implicitly on both sides through the (non-invertible~\cite{Kozik15}) functional relation $\Sigma[\Delta]$.
In an iterative scheme we obtain $\Delta^{(n+1)}=h[\Delta^{(n)}]$ as
\begin{align}
\sum\limits_{\mathbf{k}}\frac{1}{i\nu _{n}+ t _{\mathbf{k}}-\Sigma
_{\nu }^{(n)}}=\frac{1}{i\nu _{n}-\Delta^{(n+1)}_\nu-\Sigma _{\nu }^{(n)}},  \label{SC}
\end{align}%
where $\Sigma^{(n)} =\Sigma[\Delta^{(n)}]$ is considered as a functional of $\Delta^{(n)}$. Taking the
derivative of Eq.~\eqref{SC} we obtain
\begin{align}
\sum\limits_{\mathbf{k}}G_{\kv,\nu}^{2}\frac{\delta \Sigma _{\nu }}{\delta \Delta
_{\nu ^{\prime }}^{(n)}}=g_{\nu }^{2}\left( \frac{\delta \Delta _{\nu }^{(n+1)}}{%
\delta \Delta _{\nu ^{\prime }}^{(n)}}+\frac{\delta \Sigma _{\nu }}{\delta
\Delta _{\nu ^{\prime }}^{(n)}}\right), 
\end{align}
where $g_{\nu }$ is the local Green's function. From this we obtain%
\begin{align}
\mathcal{J}_{\nu \nu ^{\prime }}=\frac{\delta \Delta _{\nu }^{(n+1)}}{\delta
\Delta _{\nu ^{\prime }}^{(n)}}=\frac{1}{g_{\nu }^{2}}\left( \sum\limits_{%
\mathbf{k}}G_{\kv,\nu}^{2}-g_{\nu }^{2}\right) \frac{\delta \Sigma _{\nu }}{\delta
\Delta _{\nu ^{\prime }}}.
\end{align}%
Using the derivative of the
self-energy~\eqref{dSigma} we obtain%
\begin{align}
\mathcal{J}_{\nu \nu ^{\prime }} &=T\left( \frac{1}{g_{\nu }^{2}}\sum\limits_{\mathbf{k}}G_{\kv,\nu}^{2}-1\right)
F^{\rm loc}_{\nu \nu ^{\prime }}g_{\nu ^{\prime }}^{2}\notag\\&=\frac{1}{g_{\nu }^{2}}\mathcal{D}_{\nu
\nu ^{\prime }}g_{\nu ^{\prime }}^{2},  \label{Jak1}
\end{align}%
where 
the non-local Bethe-Salpeter kernel $\mathcal{D}$ is defined by the Equation~\eqref{DBSK} of the main text. The eigenvalues of the Jacobian \eqref{Jak1} remain smaller than one for the
physical solutions, which is provided by the eigenvalues of $\mathcal{D}_{\nu \nu
^{\prime }}$ smaller than one.

\subsection{Bethe-Salpeter equation and the non-local kernel}

The Bethe-Salpeter equation is a relation between the full vertex $F$ and the so-called irreducible vertex $\Gamma$. 
For the full local vertex $F^{\rm loc}_{\nu \nu'}$ in the charge channel (here, as well as in the main text we consider zero bosonic frequency $\omega=0$) we have the Bethe-Salpeter equation \begin{align}
\hat{F}^{\rm loc}=\left[ \hat{1}- \hat{F}^{\rm loc} \hat{x}\right] \hat{\Gamma}\label{BSeq}
\end{align}
where $\hat{x}_{\nu \nu'}=-T g_\nu^2 \delta_{\nu,\nu'}$, as in the main text. The solution of Eq.~\eqref{BSeq} is
\begin{align}
\hat{F}^{\rm loc}=\hat{\Gamma}\left[ \hat{1}+\hat{x}\hat{\Gamma}\right] ^{-1}\label{Floc}.
\end{align}

We have previously seen that $F^\text{loc}$ is the derivative of the self-energy with respect to $\Delta$, Eq.~\eqref{dSigma}.
As somewhat similar relation can be derived for the irreducible vertex 
$T \hat{\Gamma}_{\nu\nu'} = \delta \Sigma_{\nu}/\delta g_{\nu'}$.
However, in that case the derivative is done with respect to the full impurity Green's function, whereas the derivative in Eq.~\eqref{dSigma} is with respect to a bare object. The relation $\Delta\mapsto g$ is not always invertible~\cite{Kozik15} and this leads to divergences in $\Gamma$ already on the level of an impurity model, whereas $F^\text{loc}$ is divergence free (at $T>0$). By defining physical observables in terms of $F$ instead of $\Gamma$, we can avoid complications arising from these local divergences.

The generalized susceptibility in DMFT is given by \begin{align}
    \hat{\chi}^\text{DMFT} &\equiv \frac{\hat{1}}{\hat{1}+\hat{X}\hat{\Gamma}} \hat{X}. \label{eq:app:susc}
\end{align}
Note that we have defined $X$ and generalized susceptibility $\chi$ so that both are positive, after being summed over Matsubara frequencies, and our sign convention for $\Gamma$ is opposite to Georges et al.~\cite{Georges96}, who use $\chi^\text{DMFT}=\hat{1}/(\hat{1}-\hat{X}\hat{\Gamma})\hat{X}$.

Our goal is to get rid of $\Gamma$ in favor of $F$ in this expression. 
The Bethe-Salpeter Equation~\eqref{BSeq} allows us to write
\begin{align}
    \left(\hat{1}+\hat{X}\hat{\Gamma}\right)      \left(\hat{1}-\hat{x}\hat{F}\right)
  &= 1+\hat{X}\hat{\Gamma}-\hat{X}\hat{\Gamma}\hat{x}\hat{F}-\hat{x}\hat{F} \notag \\
  &= 1+\hat{X}\hat{F} - \hat{x}\hat{F} \notag \\
  &= 1-\hat{\mathcal{D}},
\end{align}
leading to the result
\begin{align}
    \hat{\chi}^\text{DMFT} &=\left(\hat{1}-\hat{x}\hat{F}\right) \frac{\hat{1}
    }{\hat{1}-\hat{\mathcal{D}}}\hat{X}.\label{eq:app:susc:df}
\end{align}
Equation~\eqref{eq:app:susc} is a geometric series consisting of repeated particle-hole scattering: $\hat{\Gamma}$ is the scattering and $\hat{X}$ is the propagation of the particle-hole pair. The latter has both a local and a non-local component, successive scattering events can occur at the same or at different sites. 
Equation~\eqref{eq:app:susc:df} is a resummation of this result, where all successive scatterings on the same site are collected in $F$. The propagatoin between scatterings is then necessarily non-local and given by $\hat{X}-\hat{x}$. Finally, the term $-\hat{x}\hat{F}$ in the prefactor ensures that processes where the first propagation is local are also included.

For the non-local vertex $F_{\bf{q},\nu \nu'}$ we find similarly to the Eq.~\eqref{Floc}%
\begin{align}
\hat{F}_{\mathbf{q}}=\hat{\Gamma}\left[
\hat{1}+\hat{X_{\bf q}}\hat{\Gamma} \right] ^{-1}.\label{EqFq}
\end{align}
This equation
can be rewritten as
\begin{align}
\hat{F}_{\mathbf{q}} &\overset{\phantom{qv=0}}{=}\hat{\Gamma}%
\left[ \hat{1}+\left(\hat{x}+\hat{\widetilde{X}}_{\bf q}\right)\hat{\Gamma}\right] ^{-1}\notag\\
&\overset{\phantom{qv=0}}{=}\hat{F}^{\rm loc}\left[ \hat{1}+\widetilde{X}_{\mathbf{q}}\hat{F}^{\rm loc}\right] ^{-1}\notag\\
&\overset{\qv=0}{=}\hat{F}^{\rm loc}\left[\hat{1}-\hat{\mathcal{D}}\right]^{-1},
\end{align}
where 
$\hat{\widetilde{X}}_{\bf q}=\hat{X}_{\bf q}-\hat{x}$
and the kernel $\mathcal{D}$ is defined by Eq.~ \eqref{DBSK}. 
We note that while the eigenvalues of $\hat{1}-\hat{\mathcal{D}}$ are positive in the stable phases, the eigenvalues of $F^{\rm loc}$, and, consequently, $F_{\bf q}$ do not have in general a definite sign, and therefore can not be used to study stability of various phases.

\subsection{Second derivative of the Landau functional}\label{app:curvature}

The first derivative of the functional $\Omega$ is given by the Eq.~\eqref{Omega1} of the main text. For the second derivative of this functional 
we obtain
\begin{align}
\frac{\delta ^{2}\Omega}{\delta \Delta _{\nu }\delta \Delta _{\nu ^{\prime }}}=
T\left(\frac{\delta g_{\nu }}{\delta \Delta _{\nu ^{\prime }}}-\frac{\delta
g_{\nu }^{\rm sc}}{\delta \Delta _{\nu ^{\prime }}}\right).
\end{align}
Differentiating the self-consistency condition $g^{\rm sc}[\Delta]=f(\Delta,g^{\rm sc}[\Delta])$, where $f(\Delta,g)$ is given by the Eq.~ \eqref{eq:fix0}, we find
\begin{align}
\frac{\delta g_{\nu }^{\mathrm{sc}}}{\delta \Delta _{\nu ^{\prime }}}
=-\sum\limits_{\mathbf{k}}G_{\kv,\nu}^{2}\left( \delta _{\nu \nu ^{\prime }}-\frac{
1}{(g_{\nu }^{\mathrm{sc}})^{2}}\frac{\delta g_{\nu }^{sc}}{\delta \Delta
_{\nu ^{\prime }}}\right). 
\end{align}
Therefore,
\begin{align}
\frac{\delta g_{\nu }^{\mathrm{sc}}}{\delta \Delta _{\nu ^{\prime }}}=-\frac{
\delta _{\nu \nu ^{\prime }}}{1-1/{(g_{\nu }^{sc})^{2}}\sum\limits_{
\mathbf{k}}G_{\kv,\nu}^{2}}\sum\limits_{\mathbf{k}}G_{\kv,\nu}^{2}.\label{dgsc}
\end{align}
Combining this with the Eq.~\eqref{dGloc}, we find at the stationary point ($g=g^{\rm sc}$)
\begin{align}
\frac{\delta ^{2}\Omega}{\delta \Delta _{\nu }\delta \Delta _{\nu ^{\prime }}} &=
\frac{T \sum_{\mathbf{k}}G_{\kv,\nu}^{2}\delta _{\nu \nu ^{\prime }}}{1-
(1/{g_{\nu }^{2}})\sum_{\mathbf{k}}G_{\kv,\nu}^{2}}+T g_{\nu
}^{2}(\delta _{\nu \nu ^{\prime }}+T F^{\rm loc}_{\nu \nu ^{\prime }}g_{\nu'}^2)\notag \\
&=\frac{T}{1-(1/g_{\nu }^{2})\sum\limits_{\mathbf{k}}G_{\kv,\nu}^{2}}\left(
\delta _{\nu \nu ^{\prime }}-\mathcal{D}_{\nu \nu'}\right) g_{\nu ^{\prime }}^{2}.\label{d2Omega}
\end{align}

Although $\Delta$ is generally a complex function, at particle-hole symmetry $\Delta$ is purely imaginary (modulo the constant Hartree contribution),
so it is sufficient to look at $\delta^2 \Omega/\delta(i\Delta)^2$. We show in the next subsection of Supplementary Material that the denominator in Eq.~\eqref{d2Omega} is positive in the region of interest and we write
\begin{align}
\frac{\delta ^{2}\Omega}{\delta i\Delta _{\nu }\delta i\Delta _{\nu ^{\prime }}} 
&=\frac{\hat{x}}{\hat{x}-\hat{X}}\left(
\delta _{\nu \nu ^{\prime }}-\mathcal{D}_{\nu \nu'}\right) \hat{x}. \label{d2OmegadiDelta2}
\end{align}
At the critical point, one eigenvalue of the Hessian $\delta^2 \Omega/\delta \Delta^2$ changes sign, the corresponding eigenvector determines the unstable direction of the free energy landscape.

Here, the leading eigenvector of the Jacobian provides the relevant direction. It is convenient to start from the right eigenvector $W_{R}$ of $\mathcal{J}$ with eigenvalue $\lambda$,
\begin{align}
    \mathcal{J}|W_{R}\rangle&=\lambda|W_{R}\rangle,
\end{align}
and the associated conjugated vector $W_{L}$ with normalization
\begin{align}
    \langle W_{L}|\mathcal{J}|W_{R}\rangle&=\lambda.
\end{align}
The right eigenvector $W_R$ of the Jacobian is related to the right eigenvector $V$ of the non-local Bethe-Salpeter kernel, with the same eigenvalue $\lambda$, in accordance with Eq.~\eqref{Jak1}. This leads to $W_{L,R}=g_\nu^{\pm 2} V$ and is consistent with the normalization $\langle V|V\rangle=1$ ($W_L$ and $W_R$ can be seen as co- and contravariant vectors in a space with metric $g_\nu^4 \delta_{\nu\nu'}$).
For the combination shown in Fig.~\ref{fig:obs} we obtain
\begin{align}
    \langle W_L| \frac{\delta^2 \Omega}{\delta (i\Delta)^2} |W_R\rangle  
&= T(\lambda-1) \langle V|\frac{g_\nu^4}{g_{\nu }^{2}-\sum\limits_{\mathbf{k}}G_{\kv,\nu}^{2}} | V\rangle,\label{WOW}
\end{align}
which goes to zero at the critical end-point, since  $\lambda \rightarrow 1$.

\subsection{Relation between different functionals}

Here we consider the reformulation of the functional originally suggested in Ref. \cite{Chitra01}, which is appropriate to establish the relation to our functional.
To this end we perform Legendre transform 
\begin{align}
\Phi _{\mathrm{imp}}[g]=\Omega _{\mathrm{imp}}[\Delta ]-T\sum_{\nu }\Delta
_{\nu }g_{\nu },
\end{align}
such that 
\begin{align}
\frac{\delta \Phi _{\mathrm{imp}}}{\delta g_{\nu }}=-T\Delta _{\nu }.
\end{align}
Furthermore, we invert the map $g^{\mathrm{sc}}[\Delta ]$ and denote its
inverse by $\Delta ^{\mathrm{sc}}[g]$. We note that while the functional $g^{%
\mathrm{sc}}[\Delta ]$ can be multivalued (see next subsection), its inverse $\Delta ^{\mathrm{sc}%
}[g]$ is well defined. Finally, we introduce the functional $\Phi \lbrack
g]=\Phi _{\mathrm{imp}}[g]+{\Phi}'[g]$, such that $\delta {{%
\Phi }'}[g]/\delta g=T \Delta _{\mathrm{sc}}[g]$. Such a functional necessarily
exists since $\Delta ^{\mathrm{sc}}[g]$ is frequency-diagonal, and can be therefore simply integrated over $g_\nu$. Therefore, we
obtain 
\begin{align}
\frac{\delta \Phi \lbrack g]}{\delta g_{\nu }}=T\left( \Delta ^{\mathrm{sc}%
}[g]-\Delta \lbrack g]\right) .\label{Eq:dPhi}
\end{align}
Differentiating once more over $g$ and using the relations \eqref{dGloc} and \eqref{dgsc}, we find 
\begin{align}
\frac{\delta ^{2}\Phi \lbrack g]}{\delta g_{\nu }\delta g_{\nu ^{\prime }}}
&=-\frac{T}{\sum_{\mathbf{k}}G_{\mathbf{k},\nu }^{2}}\left( 1-\frac{1}{g_\nu^{2}%
}\sum_{\mathbf{k}}G_{\mathbf{k},\nu }^{2}\right) +\left( \chi^{\mathrm{loc}}\right) _{\nu \nu
^{\prime }} ^{-1} \notag\\
&=-\frac{T}{\sum_{\mathbf{k}}G_{\mathbf{k},\nu }^{2}}\delta _{\nu \nu
^{\prime }}+\Gamma _{\nu \nu ^{\prime }}.  \label{d2Phi}
\end{align}
The obtained derivative~\eqref{d2Phi} is analogous to the derivative of the Baym-Kadanoff functional $\Gamma_{bk}$, discussed in Ref. \cite{Chitra01}. Using the hybridization in Eq. \eqref{Eq:dPhi} yields straightforwardly the DMFT self-consistency equation, without the need of passing to another functional, $\Gamma_{new}$ in the notation of Ref. \cite{Chitra01}.

However, the vertex $\Gamma $ may diverge at some interaction strengths in the strong coupling
regime, which makes the eigenvalues of the Eq.~ \eqref{d2Phi},
as well as the second derivative of the functional $\Gamma_\text{new}$ of Ref. \cite{Chitra01}  not positive
definite. The same concerns the functional of Ref.~\cite{Potthoff03}. To emphasize further the relation of Eq.~\eqref{d2Phi} to
our result of Eq.~\eqref{eq:OmegaStable} of the main text, we transform Eq.~\eqref{d2Phi} as 
\begin{align}
\frac{\delta ^{2}\Phi \lbrack g]}{\delta g_{\nu }\delta g_{\nu }^{\prime }}
=&\left[ \left( g_{\nu
}^{2}-\sum_{\mathbf{k}}G_{\mathbf{k},\nu }^{2}\right) \left( \delta _{\nu
\nu ^{\prime \prime }}+TF_{\nu \nu ^{\prime \prime },\omega =0}^{\mathrm{loc}%
}g_{\nu ^{\prime \prime }}^{2}\right) \right.  \notag\\
&\left. +\sum_{\mathbf{k}}G_{\mathbf{k},\nu }^{2}\delta _{\nu \nu ^{\prime
\prime }}\right] (\chi ^{\mathrm{loc}%
})^{-1}_{\nu ^{\prime \prime }\nu ^{\prime }}\frac{1}{\sum_{\mathbf{k}}G_{\mathbf{k},\nu }^{2}} \notag\\
=&\frac{1}{\sum_{\mathbf{k}}G_{\mathbf{k},\nu }^{2}}\left( \delta _{\nu \nu
^{\prime \prime }}-\mathcal{D}_{\nu \nu ^{\prime \prime }}\right) g_{\nu ^{\prime
\prime }}^{2}(\chi _{\nu ^{\prime \prime }\nu ^{\prime }}^{\mathrm{loc}%
})^{-1}.
\end{align}
We see that the considered derivative is related to the derivative of $\Omega$ in the following way
\begin{align}
\frac{\delta ^{2}\Phi \lbrack g]}{\delta g_{\nu }\delta g_{\nu }^{\prime }}=%
\frac{1}{T}\left( \frac{1}{\sum\limits_{\mathbf{k}}G_{\mathbf{k},\nu }^{2}}-%
\frac{1}{g_{\nu }^{2}}\right) \frac{\delta ^{2}\Omega \lbrack \Delta]}{\delta
\Delta _{\nu }\delta \Delta _{\nu'' }}(\chi ^{%
\mathrm{loc}})^{-1}_{\nu'' \nu ^{\prime }}.  \label{d2rel}
\end{align}
Despite the non-explicitly symmetric form of this equation, it is symmetric under $\nu \leftrightarrow \nu'$,
which can be once more verified by rewriting%
\begin{align*}
\frac{\delta ^{2}\Omega \lbrack g]}{\delta \Delta _{\nu }\delta \Delta _{\nu
^{\prime }}} 
&=
\frac{T\delta _{\nu \nu ^{\prime }}}{\left( \sum\limits_{\mathbf{k}%
}G_{\mathbf{k},\nu }^{2}\right) ^{-1}-g_{\nu }^{-2}}-T\chi _{\nu \nu ^{\prime
}}^{\mathrm{loc}},
\end{align*}
such that
\begin{align*}
&\frac{1}{T}\left[ \frac{1}{\sum\limits_{\mathbf{k}}G_{\mathbf{k},\nu }^{2}}%
-\frac{1}{g_{\nu }^{2}}\right] \frac{\delta ^{2}\Omega \lbrack g]}{\delta \Delta
_{\nu }\delta \Delta_{\nu^{\prime \prime} }}(\chi^{\mathrm{loc}} 
)^{-1}_{\nu'' \nu ^{\prime } }\\
&=(\chi^{\mathrm{loc}})^{-1}
_{\nu \nu ^{\prime }}
-\left( \frac{1}{%
\sum\limits_{\mathbf{k}}G_{\mathbf{k},\nu }^{2}}-\frac{1}{g_{\nu }^{2}}%
\right) \delta _{\nu \nu ^{\prime }}.
\end{align*}
This shows the symmetry and returns us back to the Eq.~\eqref{d2Phi}.

Let us discuss conditions of sign definiteness of the considered second derivatives. The second derivative of $\Phi$ is proportional to $(\chi _{\nu \nu ^{\prime }}^{\mathrm{loc}})^{-1},$ appearing from
the derivative $\delta \Delta /\delta g$. This makes the derivative $\delta
^{2}\Phi \lbrack g]/(\delta g_{\nu }\delta g_{\nu ^{\prime }})$ not sign
definite near MIT, which is related to multivaluedness of the map $g\mapsto
\Delta $. The multiplier $(\chi _{\nu \nu ^{\prime }}^{\mathrm{loc}})^{-1},$ is absent in the derivative of the
functional $\delta ^{2}\Omega \lbrack g]/(\delta \Delta _{\nu }\delta \Delta
_{\nu ^{\prime }})$, considered in this paper. However, this comes at the price of the new factor $[1-(1/g_{\nu }^{2})\sum\nolimits_{\mathbf{k}}G_{\mathbf{k},\nu
}^{2}]^{-1}$ in Eq.~\eqref{d2Omega}, which may not be positive definite. As
one can see from the derivation in the previous section, this factor can be traced back to
the derivative $\delta g_{sc}/\delta \Delta $. This derivative is not always positive definite,
which is in turn related to possible multivaluedness of the map $\Delta
\mapsto g_{\mathrm{sc}}$ (see next section). The advantage of using $\Omega$
is however that at strong coupling one can consider only one
relevant branch of the functional $g_{\mathrm{sc}}[\Delta ],$ as discussed
in the following section. This makes the factor $[1-(1/g_{\nu
}^{2})\sum\nolimits_{\mathbf{k}}G_{\mathbf{k},\nu }^{2}]^{-1}$ positive
definite in the strong coupling regime. In general, this reflects dichotomy
of describing the system in terms of $g$ at weak coupling vs. $\Delta $ at
strong coupling.

\subsection{Sign of the non-local bubble}

\label{app:sign:nonlocalbubble}

The sign of the non-local bubble can be ascertained by analyzing the (lattice) Green's function $G_{\kv,\nu}$ in terms of $t_\kv$ and $\Sigma_\nu$. 
Particle-hole symmetry implies that the real numbers $t_\kv$ are distributed symmetrically around zero and that $\Sigma_\nu$ is purely imaginary (after cancellation of the Hartree part with the chemical potential). To compress the notation, we introduce  $ia_\nu=i\nu-\Sigma_\nu$ ($a_\nu>0$) and write
\begin{align}
    G_{\kv,\nu} = \frac{1}{i a_\nu + t_\kv} = \frac{t_\kv - i a_\nu}{a_\nu^2+t_\kv^2}. \label{eq:app:grealim}
\end{align}
Let us rewrite the non-local bubble in terms of the density of states $\rho(\epsilon)$:
\begin{align}
    \sum_\kv G_{\kv,\nu}^2 - g_\nu ^2 
    =\int\frac{\rho(\epsilon){d\epsilon}}{(i a_\nu-\epsilon)^2}-\left(\int\frac{\rho(\epsilon){d\epsilon}}{i a_\nu-\epsilon}\right)^2.
    \label{DualBubbleDOS}
\end{align}
In the limit of large $a_\nu$ the Eq.~ \eqref{DualBubbleDOS} is proportional to the second moment of the density of states, and, therefore, positive. At small $a_\nu$ using the low energy behavior of the density of states of the square lattice, $\rho(\epsilon)\approx 2\ln(4D/|\epsilon|)/(\pi^2 D)$ 
($D=4t$ is the half bandwidth) and keeping  singular contributions, we obtain
\begin{align}
    \sum_\kv G_{\kv,\nu}^2 - g_\nu ^2 \approx 
    -\frac{2}{\pi D a_\nu}+\left(\frac{2\ln(4D/a_\nu)}{\pi D}\right)^2<0.\label{DualBubbleDOS1}
\end{align}
Numerical analysis shows that change of the sign of the dual bubble occurs at $a_\nu=a^{\rm nb}=0.1492D$. This corresponds to local Green function $g_\nu=-i g^{\rm nb}=-2.0858 i/D$.  The change of the sign of the bubble is entirely related to the van Hove singularity of the density of states, which yields the first term in the right hand side of Eq.~ \eqref{DualBubbleDOS1}, proportional to the derivative of the density of states $\rho'(a_\nu)$, 
much bigger than the second term, proportional to $\rho^2(a_\nu)$.
For smooth densities of states both $\sum_\kv G_{\kv,\nu}^2$ and $g_\nu$ remain finite for $a_\nu\rightarrow 0$ and the sign of the bubble is given by the competition of these terms. For the Bethe lattice and the simple cubic lattice we find that the bubble remains positive for all $a_\nu$. 

\begin{figure}[b]
    \centering
    \begin{center}
    \begin{tikzpicture}
      \draw (-.5,2.5) node[rotate=90]{\large$-2W\text{Im}\,g^{\rm sc}(\Delta)$};
      \draw (4,-.3) node{\large$-\text{Im}\Delta/(2W)$};
      \node[anchor=south west,inner sep=0] (image1) at
        (0,0) {\includegraphics[trim={.5cm .5cm 0 0},clip=true,width=0.4\textwidth]{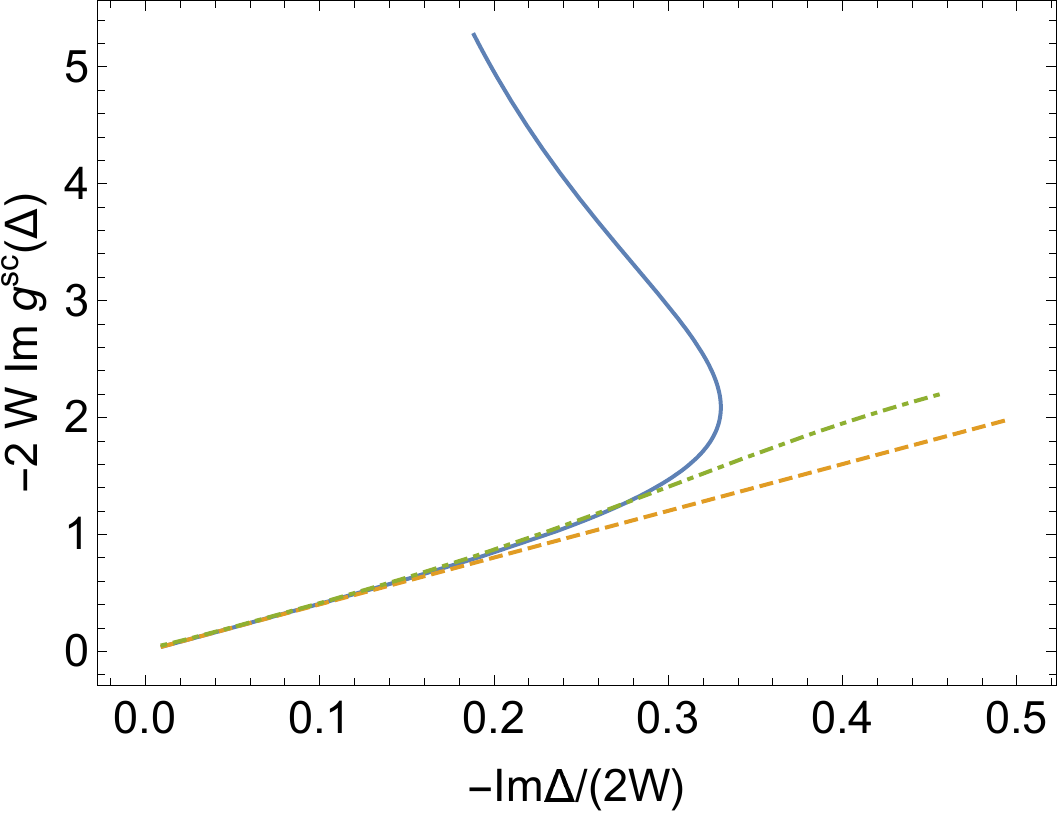}};
      \end{tikzpicture}
    \end{center}\vspace{-0.5cm}
    \caption{The function 
    $g^{\rm sc}(\Delta)$ for the square lattice (solid line, $W=D/2$), simple cubic lattice (dot-dashed line, $W=D/\sqrt{6}$), and Bethe lattice with half bandwidth $D$ (dashed line, $W=D/2$) in units of twice variance of the density of states $2W$. For the Bethe and simple cubic lattice the functions are shown in their definition domain $\Im(1/g+\Delta)>0$, while for the square lattice large $\Im g^{\rm sc}\approx 1/\Im\Delta$ are achieved at small $\Delta$. The position of the rightmost point of the solid line $(\Delta^{\rm nb},g^{\rm nb})$ corresponds to the parameters discussed in the text.
    }
    \label{fig:deltaofg}
\end{figure}

According to the relation \eqref{dgsc}, vanishing of the non-local bubble corresponds to the divergence of the derivative ${\delta g_{\nu }^{\mathrm{sc}}}/{\delta \Delta _{\nu ^{\prime }}}$. This occurs since the function $g_{\nu }^{\mathrm{sc}}(\Delta_\nu)$ (we can discuss it here as a function instead of a functional, since it is diagonal in frequency) is two-valued, see Fig.~\ref{fig:deltaofg}. The corresponding values of $g_{\nu }^{\mathrm{sc}}(\Delta_\nu)$ join each other at $|\Delta_\nu|\rightarrow \Delta^{\rm nb}-0$, where $\Delta^{\rm nb}=|i/g^{\rm nb}-i a^{\rm nb}|=0.3303 D$). The functional $\Delta^{\rm sc}[g]$, inverse to $g^{\rm sc}[\Delta]$, is also diagonal in frequency and well defined. 

In the regime of small $U$ and $T$ the corresponding values of $a_\nu$ are small at low frequencies, and the DMFT solution belongs to the "upper" branch of the function $|g^{\rm sc}(\Delta_\nu)|>g^{\rm nb}$ for these frequencies, while it belongs to the lower branch $|g^{\rm sc}(\Delta_\nu)|<g^{\rm nb}$ for higher frequencies. The increase of $U$ and/or $T$ shifts, however, the DMFT solution to the lower branch of the function $g^{\rm sc}(\Delta_\nu)$ for all Matsubara frequencies, making the corresponding functional well-defined. This also stresses universality of the Mott transition, since only one branch of the functional  $g^{\rm sc}[\Delta_\nu]$ is present for the prototypical example of Mott transition -- the Hubbard model on the Bethe lattice -- reflecting above discussed absence of a sign change of the non-local bubble for this lattice.

\end{document}